\documentclass{article}

\usepackage{authblk}
\usepackage{amsmath}
\usepackage[square, numbers]{natbib}
\usepackage[margin=0.5in]{geometry}

\usepackage{nameref}
\usepackage{url}
\usepackage{color}
\usepackage{rotating}
\usepackage[capitalise]{cleveref}

\usepackage{soul}

\usepackage{graphicx}
\graphicspath{ {./figures/} }

\newcommand\textal{\begin{adjustwidth}{1.5 em}{0 pt}}
\newcommand\mathal{\begin{adjustwidth}{2.5 em}{2.5 em}}
\newcommand\alend{\end{adjustwidth}}

\title{Powerful extragalactic jets dissipate their kinetic energy far from the central black hole}

\author[1]{Adam Leah W. Harvey*}
\author[1,2]{Markos Georganopoulos}
\author[1]{Eileen T. Meyer}
\affil[1]{University of Maryland, Baltimore County, Department of Physics, 1000 Hilltop Cir Baltimore, Maryland, USA 21230}
\affil[2]{NASA Goddard Space Flight Center}
\affil[ ]{aharvey1@umbc.edu, georgano@umbc.edu, meyer@umbc.edu}
\affil[ ]{These authors jointly supervised this work: Markos Georganopoulos, Eileen T. Meyer}

\begin{document}

\maketitle

\section{Abstract}
Accretion onto the supermassive black hole in some active galactic nuclei (AGN) drives  relativistic jets of plasma, which dissipate a significant fraction of their kinetic energy into gamma-ray radiation. The location of energy dissipation in powerful extragalactic jets is currently unknown, with implications for particle acceleration, jet formation, jet collimation, and energy dissipation. Previous studies have been unable to constrain the location between possibilities ranging from the sub-parsec-scale broad-line region to the parsec-scale molecular torus, and beyond. Here we show using a simple diagnostic that the more distant molecular torus is the dominant location for powerful jets. This diagnostic, called the seed factor, is dependent only on observable quantities, and is unique to the seed photon population at the location of gamma-ray emission. Using $62$ multiwavelength, quasi-simultaneous spectral energy distributions of gamma-ray quasars, we find a seed factor distribution which peaks at a value corresponding to the molecular torus, demonstrating that energy dissipation occurs $\sim 1$ parsec from the black hole (or $\sim 10^{4}$ Schwarzchild radii for a $10^{9}M_{\odot}$ black hole).

\section{Introduction}\label{introduction}
The spectral energy distribution (SED) of an extragalactic jet is well-described by a general two-peak structure, with a low-energy component peaking in the infrared-optical due to synchrotron emission of the relativistic particles in the jet and a high-energy component peaking in the gamma-rays due to inverse Compton scattering by the same particles \citep{BottcherGalaxies2019}. In powerful jets, the high-energy, gamma-ray component dominates over the low-energy component \citep{AckermannAPJ2015}, and is the main outlet of energy dissipation. In powerful jets, the seed photons for inverse Compton scattering are thought to be external to the jet \cite[e.g.,][]{MeyerAPJ2012, GhiselliniMNRAS2010, VercelloneAPJ2011}, and this mechanism is termed external Compton (EC) scattering. There are two primary populations of photons in powerful jet systems thought to be potential seeds for EC scattering: those from the sub-parsec-scale broad-line region, and those from the parsec-scale molecular torus \cite[e.g.,][]{SikoraApJ1994,BlazejowskiApJ2000}.

Different methods of localizing the gamma-ray emission have been proposed, with contradictory results \citep{BottcherGalaxies2019}. Short variability time-scales of gamma-ray emission in powerful jets has been used to argue that the dominant emission site must be on the sub-parsec scale, implicating the broad-line region \cite[e.g.,][]{TavecchioMNRAS2010}. The broad-line region is opaque to very-high-energy ($\geq 100$ GeV) gamma-rays, and TeV detections of powerful jets have challenged the broad-line region scenario \cite[e.g.,][]{AbramowskiAA2013, AleksicAPJL2011b}. A lack of absorption features in the total average gamma-ray spectra of the most strongly \emph{Fermi}-LAT detected powerful jets also challenges this scenario \citep{CostamanteMNRAS2018}. In contrast, several simultaneous gamma-ray/optical flares in which the optical emission exhibited polarization behavior similar to that of the very-long baseline interferometry (VLBI) radio emission implicates an emission site at or near the VLBI core, beyond the molecular torus \cite[e.g.,][]{MarscherAPJL2010}. Observations of both energy-dependent and energy-independent cooling times in gamma-ray flares of PKS 1510-089 implicate the molecular torus for some flares, and the VLBI core for other flares \citep{DotsonAPJ2015}. These prior results give contradictory answers for the site of gamma-ray emission, perhaps because they rely on single or few sources in a rare spectral state.

Here we develop and apply a diagnostic of the location of gamma-ray emission which is dependent only on long-term average observable quantities, and which can be applied to a large sample of sources, avoiding many of the problems of prior methods. We call this diagnostic quantity the seed factor, and it is unique to the seed photon population at the location of emission. Because we cannot achieve short enough integration times at all wavebands to match the variability time of extremely fast variable states, we do not apply our seed factor diagnostic to fast variability. Fitting $62$ quasi-simultaneous SEDs, we find that the molecular torus is strongly preferred as the dominant location of gamma-ray emission. This demonstrates that energy dissipation in powerful extragalactic jets occurs $\sim 1$ pc downstream of the supermassive black hole.

\section{Results}\label{results}

We have developed a diagnostic which we call the seed factor to test whether powerful extragalactic jets dissipate kinetic energy through EC in the broad-line region or the molecular torus. The derivation of the seed factor is given in the Supplementary Information. It is given by

\begin{equation}\label{sf_form}
    \text{SF}=\text{Log}10\left(\frac{U_{0}^{1/2}}{\epsilon _{0}}\right) =\text{Log}10\left(3.22\times 10^{4}\text{ }\frac{k_{1}^{1/2}\nu _{s,13}}{\nu _{c,22}} \; \text{Gauss}\right)
\end{equation}

where $U_{0}$ is the energy density of the photon population which is upscattered by the jet, in CGS units; $\epsilon _{0}$ is the characteristic (i.e., peak) photon energy of the photon population, in units of the electron rest mass; $k_{1}$ is the Compton dominance (the ratio of the inverse Compton peak luminosity to the synchrotron peak luminosity) in units of $10$; $\nu _{s,13}$ is the peak frequency of the synchrotron peak in units of $10^{13}$ Hz; $\nu _{c,22}$ is the peak frequency of the inverse Compton component in units of $10^{22}$ Hz.

The seed factor value expected due to EC scattering on a specific photon population is calculated using the energy density and characteristic photon energy of the seed photon population to be upscattered. These are known very well for both the broad-line region and the molecular torus. Since seed factor values for the molecular torus and broad-line region differ significantly, the seed factor is unique to the photon population being upscattered. The seed factor of a specific source can be determined using four observable quantities from the broadband SED: the peak frequency and peak luminosity of the synchrotron and inverse Compton emission. These parameters are purely observationally derived and can be constrained well using existing data. For these reasons the seed factor diagnostic is very robust. After calculating the seed factor for a source, it can be compared to the expected value for either the broad-line region or molecular torus. Emission mechanisms other than external Compton scattering can also be tested against, since in such a case there is no \emph{a priori} reason that the seed factor should be a single value for all sources or all emission states.

To calculate the seed factor expected for the broad-line region and the molecular torus (full details in Supplementary Information), an estimate of the energy density and characteristic photon energy of each photon population are needed. Reverberation mapping and near-infrared interferometric studies of radio-quiet AGN \cite[e.g.,][]{BentzAPJ2013, SuganumaAPJ2006, Pozo_NunezAA2014, KishimotoAA2011} combined with studies of the covering factors of the broad-line region \citep{GhiselliniMNRAS2009} and molecular torus \citep{HaoAPJ2010, HaoAPJ2011, MaiolinoAA2007} imply that both the broad-line region and molecular torus have a constant energy density across sources. Combined with estimates of the characteristic photon energy of the broad-line region \citep{MalmroseAPJ2011} and the molecular torus \citep{TavecchioMNRAS2008}, the expected seed factor for the broad-line region and the molecular torus are, respectively,

\begin{eqnarray}
    \text{SF}_{\text{BLR}} = 3.29 \pm 0.11\\
    \text{SF}_{\text{MT}}  = 3.92 \pm 0.11
\end{eqnarray}

\begin{figure}
    \centering
    \includegraphics[width=\textwidth]{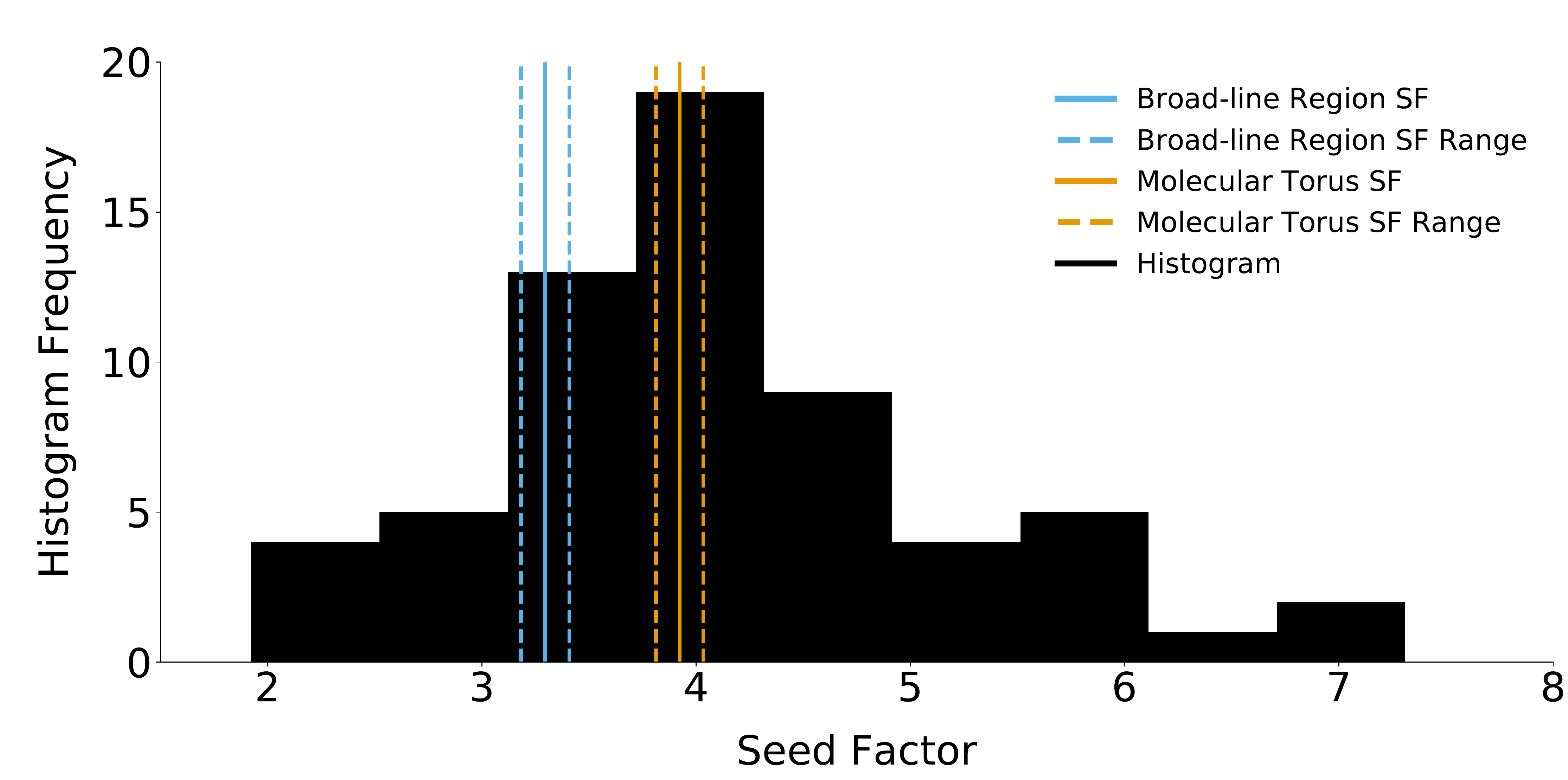}
    \caption{\textbf{Histogram of FSRQ seed factors.} A histogram showing the distribution of seed factors for our sample of FSRQs. The histogram is plotted in black. The expected broad-line region seed factor and $1\sigma $ confidence interval are shown in blue, with solid and dashed lines, respectively, on the left. The expected molecular torus values are plotted similarly, on the right, in orange. See Sections 1.2 and 1.3 of the Supplementary Information for information on calculation of the plotted confidence intervals. The histogram was binned using the \texttt{auto} option for the binning in \texttt{matplotlib.pyplot.hist}. The histogram peaks at about the expected seed factor value of the molecular torus.}
    \label{fsrq_histogram}
\end{figure}

\begin{figure}
    \includegraphics[width=\textwidth]{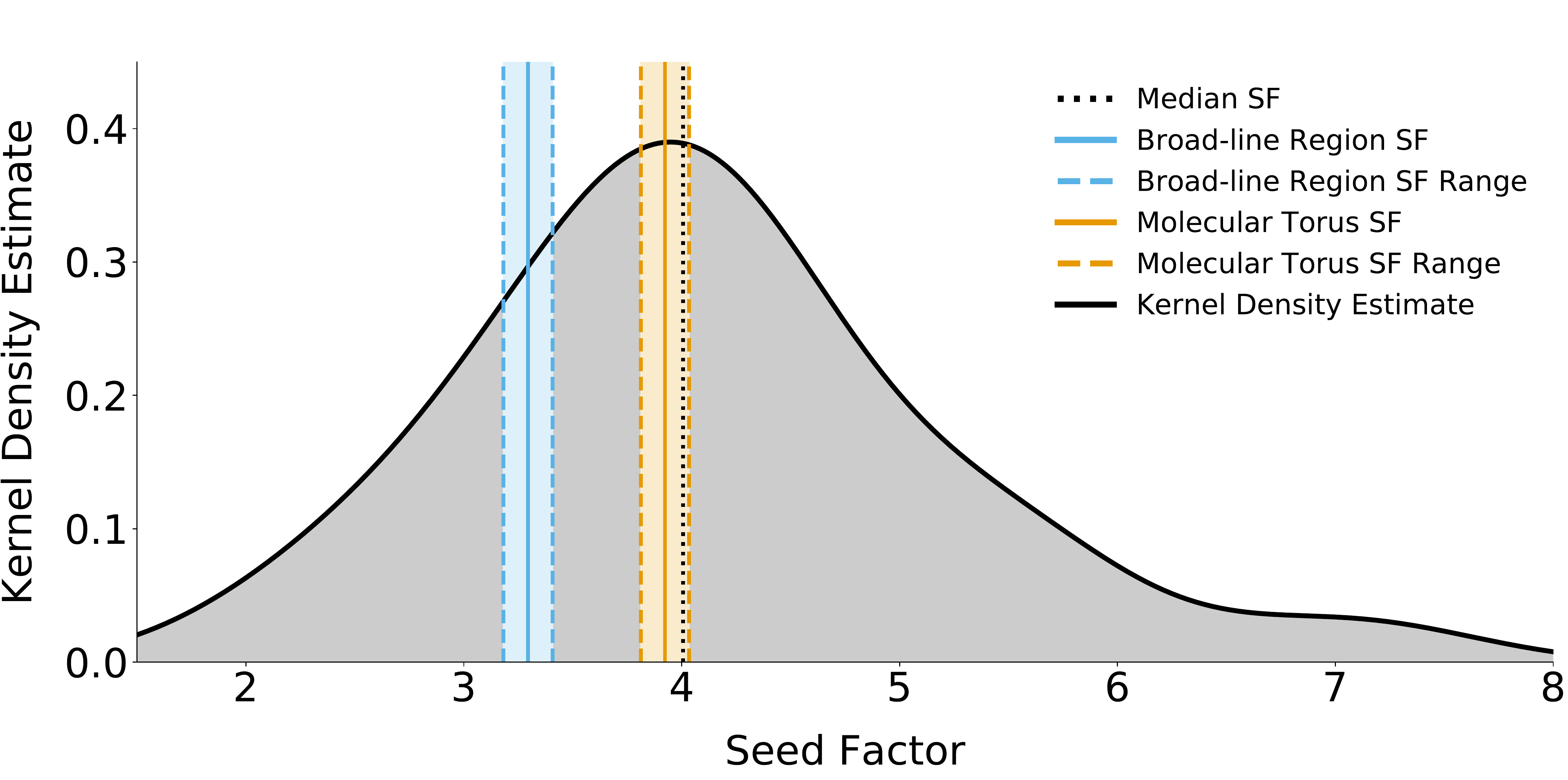}
    \caption{\textbf{Kernel density estimate of FSRQ seed factors.} A kernel density estimate (KDE) of the seed factor distribution of the combined sample of $62$ SEDs (see Sections 2.3 and 2.5 of the Supplementary Information for more details). The KDE is plotted as a thick black line, with the area underneath shaded in gray as a visual aid.  The median of the distribution is denoted by a black dotted line. The expected broad-line region seed factor and $1\sigma $ confidence interval are shown in blue, with solid and dashed lines, respectively, on the left. The expected molecular torus values are plotted similarly, on the right, in orange. See Sections 1.2 and 1.3 of the Supplementary Information for information on calculation of the plotted confidence intervals. The KDE peaks within the $1\sigma $ confidence interval of the molecular torus, nearly coincident with the expected value of the molecular torus. The KDE has the general appearance of a normal distribution. A bootstrapped (using $10^{7}$ variates) two-sided Kolmogorov-Smirnov test indicates that normality cannot be rejected significantly (rejection significance of $1.42\sigma$; p-value $0.16$).}
    \label{kde_fig}
    \centering
\end{figure}

To investigate if there is a population trend in the location of energy dissipation, we study the distribution of the seed factor. The histogram of seed factors is plotted in Figure \ref{fsrq_histogram}. However, a histogram does not give any direct information about measurement uncertainties. To better visualize the measured seed factor distribution, we therefore implemented a kernel density estimation which incorporates the uncertainties via bootstrapping (see Sections 2.3 and 2.5 of the Supplementary Information). The kernel density estimate we have implemented uses Silverman's Rule \citep{Silverman_density_estimation_textbook_1998} for the bandwidth to produce a reasonably unbiased smoothed kernel density estimate of the seed factor distribution (see Figure \ref{kde_fig}). As can be seen from this kernel density estimate, the seed factor distribution peaks within the $1\sigma $ confidence interval of the molecular torus seed factor.

We further calculated the median of the seed factor distribution, and calculated the $1\sigma $ confidence interval for the median via bootstrapping (using $8\times 10^{5}$ variates). The median seed factor is

\begin{equation}
    \text{SF}_{\text{median}} = 4.01^{+0.10}_{-0.12}
\end{equation}

We tested the seed factor distribution for normality and compared the distribution with that measured for weak extragalactic jets (which are known to emit through processes other than external Compton scattering) to test for consistency with external Compton scattering. Finding a rejection significance of normality of $1.42\sigma $ (using a bootstrapped two-sided Kolmogorov-Smirnov test) and finding that the weak and powerful jet distributions are statistically different, the seed factor distribution of powerful jets is consistent with external Compton scattering.

To test whether the seed factor distribution is inconsistent with emission from either the broad-line region or the molecular torus, we calculate the rejection significance of the expected seed factors. Using a bootstrapping method, we calculated the significance of rejecting that the median of the observed distribution is different from the respective expected values. The rejection significances calculated, in terms of standard deviations, for the broad-line region and the molecular torus, respectively, are,

\begin{eqnarray}
    \sigma(\text{Obs}_{\text{median}}-\text{BLR}_{\text{median}}) = 6.10\sigma \\
    \sigma(\text{Obs}_{\text{median}}-\text{MT}_{\text{median}}) = 0.71\sigma
\end{eqnarray}

We therefore conclude that, except possibly in periods of fast variability, the molecular torus is strongly preferred as the single dominant location of energy dissipation in powerful extragalactic jets.

\section{Discussion}\label{sec:discussion}
Our finding sets specific constraints on jet models: there is no substantial steady-state jet energy dissipation at scales less than $\sim 1$ pc. Within this distance the flow has to collimate and achieve an opening angle of a few degrees, and at the same time accelerate to bulk Lorentz factors of 10 to 50, as required by VLBI studies \citep{JorstadAJ2005}. Major particle acceleration and subsequent dissipation of the order of $10\%$ of the jet power \citep{NemmenSci2012} must take place beyond the sub-pc broad-line region and within the $\sim$ pc scale molecular torus. This conclusion does not rest on any single source, but rather on clear observables for an entire population of powerful jets. These observables are measured using quasi-simultaneous multi-wavelength SEDs, which act as snapshots of the broadband emission of each source, reducing biases due to averaging. Quasi-simultaneous SEDs reduce the chance of inter-band integration mismatches. That is to say that a quasi-simultaneous SED is unlikely to, for example, contain X-ray data taken during a high state while other data is taken during a low state. This work can be expanded as additional well-sampled SEDs become available.

We note here that the seed factor derived from a single SED of a single source would be reliable if we had near-perfect knowledge of the SED at all frequencies over exactly matching timeframes. This is essentially never the case. It is, however, possible to localise the emission for a single source with a large number of SEDs of that source.

A more detailed understanding of the broad-line region and the molecular torus would enable use of the seed factor to localize the site of energy dissipation more precisely, and to study the possible distribution of dissipation sites across both time and sources. The broad-line region and the molecular torus are expected both to be stratified \cite[e.g.,][]{FinkeAPJ2016, AbolmasovMNRAS2017} and possibly exist as components of an accretion disk wind \cite[e.g.,][]{BoulaMNRAS2019}. Furthermore, minor deviations from the scaling relations we have used are currently under debate \cite[e.g.,][]{KishimotoAA2011, BurtscherAA2013, BentzAPJ2013, GrierAPJ2017, DuAPJ2018}. The magnitude and nature of the effects of such a possible deviation would have implications for a more detailed analysis of the seed factor distribution. Conversely, the detailed structure of AGN hosting powerful jets can be constrained by requiring models to produce a seed factor distribution consistent with that observed.

Our result also argues that the VLBI core  is not the dominant location of gamma-ray emission. The fact that the seed factor distribution has a peak and this peak is around the molecular torus seed factor would be a very unlikely coincidence if the emission location was typically the VLBI core.

\section{Methods}\label{methods}
    \subsection{Description of SED Samples}
    We used $4$ samples (as described in Section 1.4 of the Supplementary Information) of well-sampled quasi-simultaneous, multiwavelength SEDs of powerful blazars (flat spectrum radio quasars, FSRQS) to calculate the seed factor for a representative population of sources. We fit the published SEDs to obtain the peak frequencies and luminosities (these values are given in Supplementary Tables 2-6). To reliably constrain the peak luminosity and the peak frequency of the synchrotron and inverse Compton components, these SEDs needed to be well-sampled in frequency (see Section 2.2 of the Supplementary Information for information on testing spectral coverage). Quasi-simultaneity is required due to the fact that blazars are variable sources, and we thus require the ability to reliably fit individual states of emission. Two of these samples are catalog samples from the literature, with SEDs published in their respective papers. One of these two samples is made up of $5$ different sub-samples selected based on different criteria in a project of the Planck Collaboration \citep{GiommiAAP2012, Planck_CollaborationAAP2011}. The second is the Fermi LAT Bright AGN Sample \citep{AbdoAPJ2010v}. The third is from a sample of blazar SEDs for some sources observed in the Tracking Active Galactic Nuclei with Austral Milliarcsecond Interferometry program (TANAMI; \citep{OjhaAAP2010}). The fourth sample is a sample of SEDs taken from a literature search of two different well-observed sources ($3\text{C } 279$ and $3\text{C } 454.3$; see Section 1.4.4 of the Supplementary Information for more information). Some SEDs were excluded on the basis of poor coverage and scattering outside of the Thompson regime (that is, if $\nu _{c}>10^{24}$ Hz; see Section 1.4 and 2.2 of the Supplementary Information for details on SED exclusions). In the case of SEDs from our literature search, SEDs were also excluded in the case of either non-quasi-simultaneous SEDs or observations of non-steady states (i.e., when there was clear variability reported for the time of the observations; the seed factor is applicable only to steady-state emission).

    \subsection{SED Fitting}
    We fit the obtained SEDs using maximum likelihood regression implemented through a simulated annealing optimization algorithm. Errors were calculated using the likelihood ratio method (i.e., Wilk's Theorem), implemented through a combination of bootstrapping, kernel density estimation, and polynomial interpolation (as described in the Supplementary Information). There are $62$ SEDs in total which were useable based on our criteria of quasi-simultaneity, coverage, steady state emission, and Thomson regime scattering. We calculated the seed factor and $1\sigma $ uncertainties for these $62$ SEDs.

    The peak of any individual SED is only as reliable as the model and the data. The formal errors on the fits from application of Wilk's Theorem implicitly do not take into account the error from variability and the fact that the spectra are not intrinsically perfect polynomials but only well-approximated as such. We assume that the extra error contribution averages out in the sample estimate of where the seed factor distribution peaks. This is the main reason that we do not attempt to make a claim about the location in any single source, and instead use many measurements to try to find the ensemble answer, which implicitly relies on the errors averaging out. This approach uses the population to provide statistical constraints on the behavior of powerful jets as a source class.

    \subsection{Statistical Analysis of the Seed Factor Distribution}
    We used a two-sided bootstrapped Kolmogorov-Smirnov test \citep{FeigelsonTextbook2012} to test the normality of the distribution, using a rejection significance threshold of $2\sigma $. Normality of the seed factor distribution cannot be rejected significantly (rejection significance of $1.42\sigma $; p-value of $0.16$).

    Consistency with external Compton scattering was tested by comparing the seed factor distribution of powerful extragalactic jets with that of weak extragalactic jets, which are known to emit through processes other than external Compton (see Section 4 of the Supplementary Information for more information). The distribution for powerful jets was found to be different than that for weak jets, further implying an external Compton origin of the emission from powerful jets. Due to the normality, sharp peak, and divergence from weak extragalactic jets, the observed seed factor distribution is consistent with a single, dominant location of energy dissipation in powerful extragalactic jets.

    To test whether the seed factor distribution is inconsistent with either the broad-line region or the molecular torus, we calculated the significance of rejection of the respective expected seed factors. We implemented a bootstrapping procedure to produce a distribution of the difference between the median of the observed seed factor distribution and the expected seed factors. This was done by performing a large sample-size bootstrap on the seed factor distribution of the $62$ SEDs, and creating a proxy sample of the same size for the broad-line region and molecular torus, drawing, for each, $62$ random samples from  their respective normal distributions (as defined by the calculated values and uncertainties given earlier). For each triplet of medians, we calculated the difference between the bootstrapped observed seed factor median and the proxy sampled broad-line region and molecular torus seed factor medians. Each distribution of differences is consistent with being normally distributed, as determined by a two-sided Kolmogorov-Smirnov test (using a rejection significance threshold of $2\sigma $, normality is rejected at the $1.75\sigma $ (p-value $0.08$) and $1.82\sigma $ (p-value $0.07$) level for the broad-line region and molecular torus difference distributions, respectively). Given the normality of these distributions, the significance of rejecting the hypothesis that the medians are the same was then calculated as the absolute value of the ratio of the mean and standard deviation of each sample of differences (i.e., the number of standard deviations between zero and each median). The rejection significances of the seed factor differences, in terms of standard deviations, for the broad-line region and molecular torus respectively, are,

    \begin{eqnarray}
        \sigma(\text{Obs\textsubscript{median}-BLR\textsubscript{median}}) = 6.10\sigma \\
        \sigma(\text{Obs\textsubscript{median}-MT\textsubscript{median}}) = 0.71\sigma
    \end{eqnarray}

\section{Data Availability}
All data used is publicly available. Data available as machine-readable tables can be found using the following DOI names: LBAS: 10.26093/cds/vizier.17160030 [\url{https://doi.org//10.26093/cds/vizier.17160030}]; Giommi: 10.26093/cds/vizier.35410160 [\url{https://doi.org/10.26093/cds/vizier.35410160}]; DSSB: 10.26093/cds/vizier.35910130 [\url{https://doi.org/10.26093/cds/vizier.35910130}]. Some data were not available as machine-readable tables, but were extracted from plots (using Dexter \citep{DemleitnerDexter2001}) or tables in published papers, as noted in the Supplementary Information. The digitized tables are available upon request.

\section{Code Availability}
Results can be reproduced using standard free analysis packages. Methods are fully described. Code can be made available by request. Version and accession information for publicly available software packages referenced are as follows: matplotlib.pyplot.hist (matplotlib 2.2.3) [\url{https://matplotlib.org/2.2.3/api/_as_gen/matplotlib.pyplot.hist.html#matplotlib.pyplot.hist}]; scipy.interpolate.splrep (scipy 1.2.1) [\url{https://docs.scipy.org/doc/scipy/reference/generated/scipy.interpolate.splrep.html#scipy.interpolate.splrep}]; scipy.interpolate.sproot (scipy 1.2.1) [\url{https://docs.scipy.org/doc/scipy/reference/generated/scipy.interpolate.sproot.html#scipy.interpolate.sproot}]; sklearn.neighbors.KernelDensity (scikit-learn 0.19.2) [\url{https://scikit-learn.org/0.19/modules/generated/sklearn.neighbors.KernelDensity.html#sklearn.neighbors.KernelDensity}]; gammapy (gammapy 0.9) [\url{https://docs.gammapy.org/0.9/}]; Dexter (Dexter 0.5a) [\url{http://dexter.sourceforge.net/}]; Simulated Annealing (Simulated Annealing 3.2) [\url{https://eml.berkeley.edu/Software/abstracts/goffe895.html}].

\section{Acknowledgements}
A.L.W.H. and M.G. acknowledge support from \emph{Fermi} grant NNX14AQ71G.

\section{Author Contributions}
A.L.W.H. collected all data from the literature and developed and performed all relevant analyses. M.G. and E.T.M. supervised the work.

\section{Competing Interests}
The authors declare no competing interests.

 \newcommand{\noop}[1]{}

\end{document}


\maketitle

\section{Supplementary Information}\label{SI}
\subsection{Derivation of the Seed Factor} \label{sec:SF}
    We assume here that gamma-ray emission is due to external Compton scattering. We begin with the peak energies (in units of electron rest mass energy) of the synchrotron and inverse Compton SED components, as seen in the observer's frame. These are, respectively,

    \begin{eqnarray}
    \epsilon_{s} = \frac{B}{B_{cr}} \gamma^{2}\delta/(1+z)\label{sync_peak} \\
    \epsilon_{c} = \frac{4}{3} \epsilon _{0}\gamma^{2}\delta^{2}/(1+z) \label{IC_peak}
    \end{eqnarray}

    where $\epsilon _{0}$ is the characteristic energy of the external seed photons in units of the electron rest mass energy, $B$ is the magnetic field in the jet,  $\delta $ is the Doppler factor, $\gamma$ is the Lorentz factor of the electrons responsible for the SED peak energy emission of both the synchrotron  and inverse Compton SED components, and $B_{cr}$ is the critical magnetic field ($B_{cr}=m_{e}^{2}c^{3}/e\hbar $).

   Supplementary Equation (\ref{IC_peak}) is valid if the electron scattering takes place in the Thomson regime, which holds if,
    \begin{equation}\label{Thomson_criterion}
    \nu_c \underset{\sim }{<} 3.5 \times 10^{23}/(1+z) \;{\rm Hz}.
    \end{equation}
    Here, as in the rest of the paper we have assumed that
    the highest energy possible for external seed photons in the broad-line region and molecular torus are UV emission-line photons which have $\epsilon _{0}\approx 3 \times
    10^{-5}$ \citep{TavecchioMNRAS2008}.
    For powerful blazars, generally, $\langle \nu _{c}\rangle \approx 10^{22}$ Hz. Thus the peak of the inverse Compton component is generally produced in the Thomson regime for powerful blazars.

    Dividing Supplementary Equation (\ref{IC_peak}) by Supplementary Equation (\ref{sync_peak}) and solving for $B/\delta $,

    \begin{equation}\label{bd_scattering}
    \frac{B}{\delta } = \frac{4\epsilon _{0}\epsilon _{s}B_{cr}}{3\epsilon _{c}}.
    \end{equation}

   We now consider the Compton dominance. This is the ratio of the observed peak inverse Compton luminosity, given by

   \begin{equation}
        L_{c}=\frac{16}{9} \sigma_{T} c \beta_p \gamma_{p}^{2} n(\gamma_{p})U_{0} \delta^{6}
   \end{equation}

  and the observed synchrotron peak luminosity, given by

    \begin{equation}
        L_{s}=\frac{4}{3}\sigma_{T} c \beta_p \gamma_{p}^{2} n(\gamma_{p})U_{B}\; \delta^{4}
    \end{equation}

    Here $\sigma_T$ is the Thomson cross section, $c$ is the speed of light, $\gamma_p$ is the Lorentz factor of the electrons responsible for the peaks of the synchrotron and inverse Compton components, $n(\gamma_p)$ is the electron density distribution at $\gamma_p$, $\beta_p$ is the  speed of these electrons in units of $c$,  $\Gamma$ is the  bulk Lorentz factor, $U_0$ is the external photon field energy density in the galaxy frame, and $U_B=B^2/8\pi$ is the magnetic field energy density.
Their ratio is given by
    \begin{equation}\label{Compton_dominance}
    k = \frac{L_{c}}{L_{s}} = \frac{32\pi \delta ^{2}U_{0}}{3B^{2}},
    \end{equation}
    where $U_0$ is the energy density of the external photon field.

    Solving for $B/\delta $, equating to Supplementary Equation (\ref{bd_scattering}), and doing some simple algebra, we find
    \begin{equation}\label{seed_factor_exp}
    \frac{U_{0}^{1/2}}{\epsilon _{0}} =3.22\times 10^{4}\text{ }\frac{k_{1}^{1/2}\nu _{s,13}}{\nu _{c,22}} \; \text{Gauss},
    \end{equation}

    where $k_{1}$ is the Compton dominance in units of 10, $\nu _{s,13}$ is the synchrotron peak in units of $10^{13}$ Hz, and $\nu _{c,22}$ is the EC peak in units of $10^{22}$ Hz.

    This gives us the seed factor,

    \begin{equation}\label{seed_factor}
        \text{SF}=\text{Log}10\left(\frac{U_{0}^{1/2}}{\epsilon _{0}}\right) =\text{Log}10\left(3.22\times 10^{4}\text{ }\frac{k_{1}^{1/2}\nu _{s,13}}{\nu _{c,22}} \; \text{Gauss}\right).
    \end{equation}

    We note here that similar equations have been presented in \cite{SikoraApJ2009} without discussing how these equations can be used to infer the precise location of gamma-ray emission in FSRQs.

    \subsection{Estimating the Seed Factor in the Molecular Torus} To calculate the expected seed factor of the molecular torus, we need an estimate of its energy density and characteristic photon energy. In the molecular torus, reverberation mapping \cite[e.g.,][]{SuganumaAPJ2006, Pozo_NunezAA2014} and near-infrared interferometric studies \citep{KishimotoAA2011} of radio-quiet sources find that the inner radius of the molecular torus scales as $L_{d,45}^{1/2}$ (where $L_{d,45}$ is the accretion disk luminosity in units of $10^{45}$ erg s$^{-1}$). We assume this relation holds for radio-loud AGN. This relation implies that the molecular torus energy density is constant with both molecular torus size and luminosity. The best-fit relation and uncertainties on the scaling factor were found using the data in \cite{SuganumaAPJ2006} and uncertainties were calculated using Wilk's Theorem. For the molecular torus we find $R_{\text{MT}} = 4.83\pm 0.23 \times 10^{18}L_{d,45}^{1/2}$. The fraction of $L_{d}$ which is reprocessed by the molecular torus (the covering factor) is $\xi _{\text{MT}} \sim 0.2$ \citep{HaoAPJ2010, HaoAPJ2011, MaiolinoAA2007}. We estimate the uncertainty on the covering factor as $50\%$ of the expected value. This is justified by the distribution of molecular torus covering factors for high-$z$, luminous quasars found in \citep{MaiolinoAA2007}. We therefore estimate the molecular torus covering factor as $\xi _{\text{MT}}=0.2\pm 0.1$. The energy density of the molecular torus is therefore $U_{0,\text{MT}}= 2.28\pm 1.16\times 10^{-5}$ erg cm\textsuperscript{-3}. The molecular torus spectrum can be approximated using a blackbody spectrum with a temperature of $T=1200$ K \citep{MalmroseAPJ2011}. This gives a characteristic photon energy for the molecular torus of $\epsilon _{0,\text{MT}} = 5.7\times 10^{-7}$, in units of electron rest mass energy. These quantities imply a molecular torus seed factor of $\text{SF}_{\text{MT}} = 3.92\pm 0.11$.

    \subsection{Estimating the Seed Factor in the Broad-Line Region} Reverberation mapping of radio-quiet active galactic nuclei finds that the radius of the broad-line region is $R_{\text{BLR}} = 2.76\pm 0.20 \times 10^{17}L_{d,45}^{1/2}$ \citep{BentzAPJ2013}. We assume that this holds for radio-loud sources. The covering factor of the broad-line region is $\xi _{\text{BLR}}\sim 0.1$ \citep{GhiselliniMNRAS2009}. Because the molecular torus is expected to be the same material as the broad-line region, but beyond the sublimation radius \citep{NenkovaAPJ2008}, the uncertainty in the broad-line region covering factor should be approximately the same as for the molecular torus. We therefore assume a $50\%$ uncertainty in the broad-line region covering factor. The broad-line region covering factor is therefore estimated as $\xi _{\text{BLR}}=0.1\pm 0.05$. With this scaling relation and covering factor, the energy density of the broad-line region is $U_{0,\text{BLR}}= 3.49 \pm 1.80\times 10^{-3}$ erg cm\textsuperscript{-3}. The broad-line region SED can be approximated by a blackbody with peak frequency of $\nu _{0} = 1.5\nu _{\text{Ly}_{\alpha}}$ \citep{TavecchioMNRAS2008}. The characteristic photon energy of the broad-line region is therefore  $\epsilon _{0,\text{BLR}} = 3\times 10^{-5}$, in units of electron rest mass energy. These assumptions imply a broad-line region seed factor of $\text{SF}_{\text{BLR}} = 3.29\pm 0.11$.

\subsection{Description of SED Samples}
    Plots of the SEDs used and their fits are included in Supplementary Fig. \ref{giommi_postage}-\ref{lss_postage}.

    \subsubsection{LBAS Sample} The LBAS is the Fermi LAT Bright AGN Sample. A subsample of $48$ quasi-simultaneous, multiwavelength SEDs for sources in the LBAS were presented in \citep{AbdoAPJ2010v}, selected from the LBAS solely on availability of Swift observations carried out between $2008$ May and $2009$ January. This selection criterion implies that the LBAS subsample should be relatively unbiased in terms of selection effects. The distribution of redshifts, optical, X-ray, and gamma-ray fluxes of the LBAS Sample are consistent with being the same as for the full LBAS \citep{AbdoAPJ2010v}. Data for this sample is available via public repository at Vizier \cite{LBAS_Vizier_Data}. However, due to evident data missing in the tables available at Vizier, all data was extracted from plots in \citep{AbdoAPJ2010v} using Dexter \citep{DemleitnerDexter2001}.

    \subsubsection{Giommi Sample} Part of the Giommi Sample is a sample of $105$ quasi-simultaneous, multiwavelength SEDs belonging to $3$ flux limited samples based on Planck, Swift, and Fermi data respectively as presented in \citep{GiommiAAP2012}. The other part of the Giommi Sample is taken from a companion paper from the Planck Collaboration \citep{Planck_CollaborationAAP2011} which included $28$ quasi-simultaneous SEDs which were not represented in \citep{GiommiAAP2012} (i.e., not overlapping with the sample in \citep{GiommiAAP2012}). Due to the companion nature of the Planck Collaboration paper, we have decided to refer to this collection of $2$ samples as a single sample (we do, however, include the data for the Planck Collaboration paper in its own table). Swift-UVOT and \emph{Fermi}-LAT data from \citep{GiommiAAP2012} was accessed via public repository at Vizier \cite{Giommi_Vizier_Data}. Due to evident inconsistencies found in the tables available at Vizier, radio, ground-based optical, and Swift-XRT data from \citep{GiommiAAP2012} was extracted from plots in the cited paper using Dexter \citep{DemleitnerDexter2001}. All data from \citep{Planck_CollaborationAAP2011} was extracted from plots in the cited paper using Dexter \citep{DemleitnerDexter2001}.

    \subsubsection{DSSB Sample} The DSSB Sample is the sample of SEDs in \citep{KraussAA2016}. DSSB stands for Dynamic SEDs of Southern Blazars. This is a sample of SEDs of the $22$ gamma-ray brightest sources (according to the Fermi-LAT Third Source Catalog; \citep{Acero2015}) observed in the Tracking Active Galactic Nuclei with Austral Milliarcsecond Interferometry program (TANAMI; \citep{OjhaAAP2010}). Time periods of steady activity were identified in \citep{KraussAA2016} by using Bayesian blocks to determine periods of constant gamma-ray flux, being subdivided further when such periods were longer than a year. \citep{KraussAA2016} then searched for observations in other wavebands which were quasi-simultaneous to these identified time-periods. Data for this sample was accessed via public repository at Vizier \cite{DSSB_Vizier_Data}.

    We found that the DSSB Sample was needed to include enough BL Lacertae object SEDs for our testing against weak extragalactic jets as described in Section \ref{SI:weak_jets} of the Supplementary Information. The DSSB contributes somewhat minimally to our sample of FSRQs (only $11$ FSRQ SEDs from the DSSB Sample are included in our final sample of SEDs).

    \subsubsection{Literature Search Sample} The Literature Search Sample is a sample of quasi-simultaneous, multiwavelength SEDs taken from the literature. Initially $7$ sources were included in this sample: 3C 279, 3C 454.3, 3C 345, PKS 1510-089, PKS 0528+134, PKS 0420-015, and 1502+106. SEDs were excluded from the final sample based only on simultaneity, wavelength coverage, and steady-statedness. This left only SEDs from 3C 279 and 3C 454.3 in the sample. The SEDs included in the Literature Search Sample for these two sources (for completeness, including those which did not pass the cut on peak Compton frequency) are in \citep{WehrleAPJ1998, AbdoAPJ2009, VercelloneAPJ2010, VercelloneAPJ2011, HayashidaAPJ2015, HayashidaAPJ2012, PaccianiAPJ2010}. All data in this sample was extracted from either tables or plots in the cited papers. Data was extracted from plots, using Dexter \citep{DemleitnerDexter2001}, in cases of evident inconsistencies in tables, missing data in tables, or in cases for which tables were not provided.

    \begin{figure}[h]
        \centering
        \includegraphics[width=\textwidth]{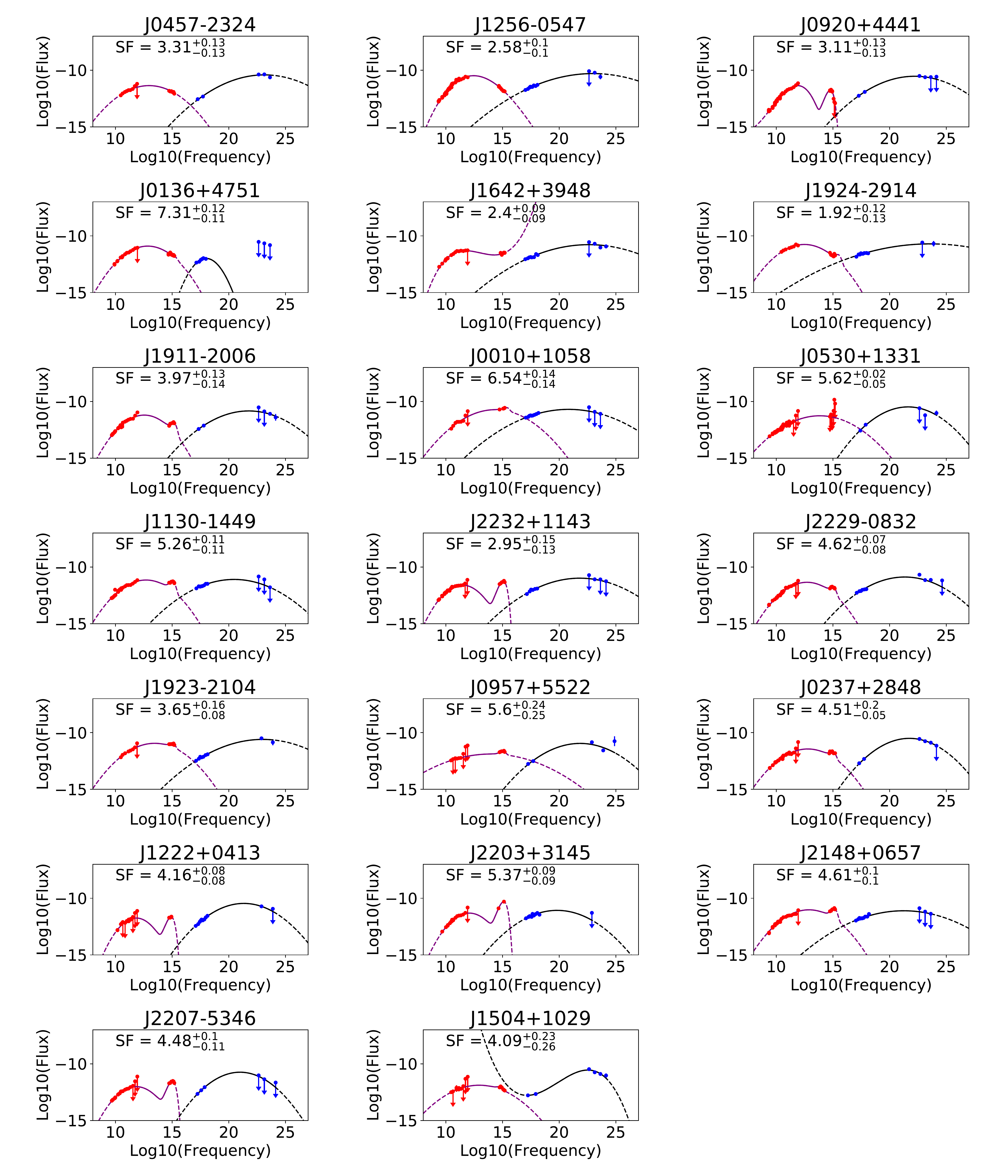}
        \caption{\textbf{Giommi Sample FSRQ SEDs.} The SEDs for the FSRQs in the Giommi Sample which passed all selection criteria. Red data points represent data we identified as being attributable to the synchrotron peak, big blue bump emission, or extended emission. Blue data represent data we identified as being primarily attributable to inverse Compton emission. The green curve shows the fit to the red data; the solid parts show the portion within the range of the red data, while the dashed parts show the extension of the fit beyond the range of the red data. The black curve represents the same, but for the blue data. Frequencies are in Hertz. Fluxes are in erg cm\textsuperscript{-2} s\textsuperscript{-1}. Error bars are as they were presented by the original authors. Downward pointing arrows represent upper limits as they were presented by the original authors.}
        \label{giommi_postage}
    \end{figure}

    \begin{figure}[h]
        \centering
        \includegraphics[width=\textwidth]{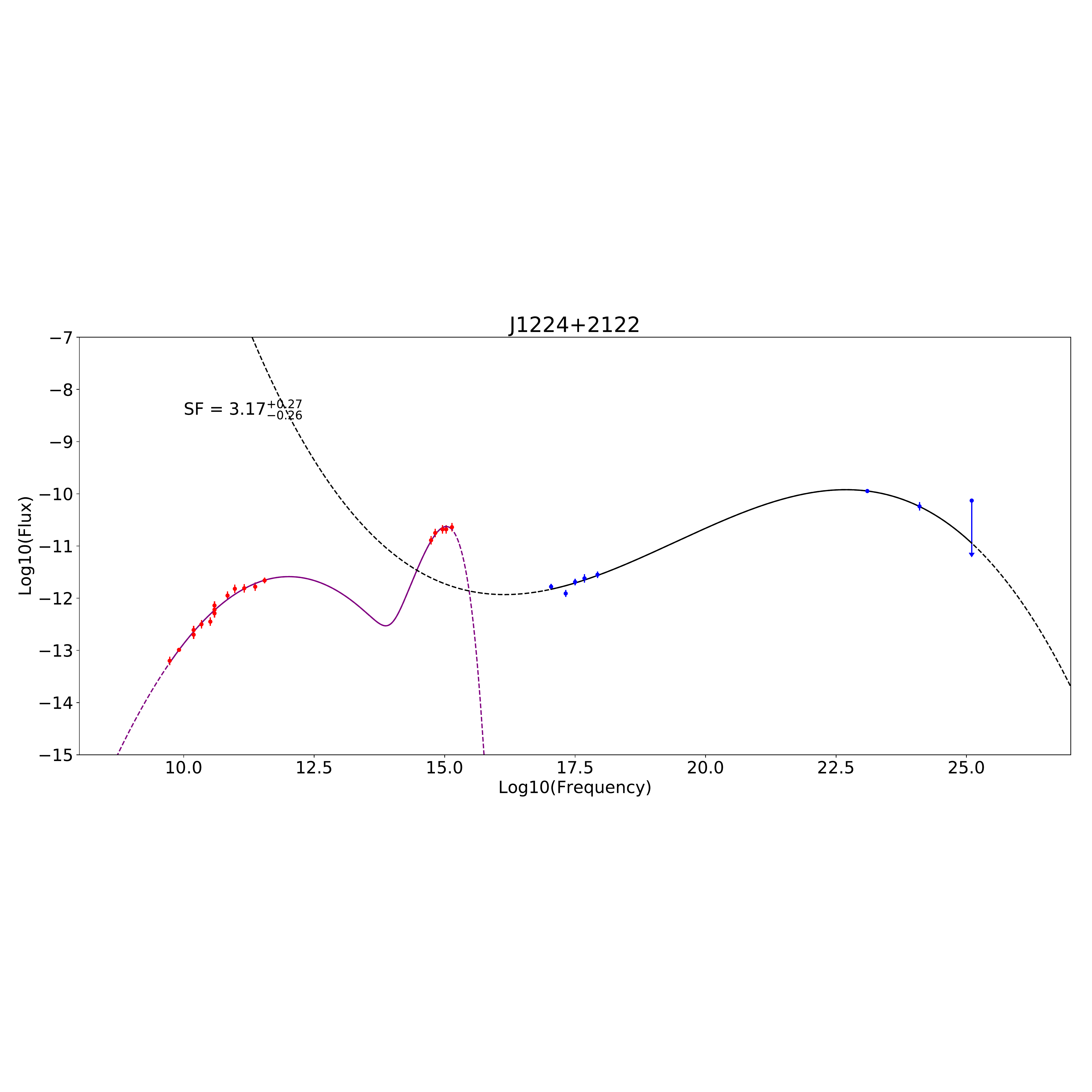}
        \caption{\textbf{Planck Sample FSRQ SEDs.} The SEDs for the FSRQs in the Planck Sample which passed all selection criteria. Red data points represent data we identified as being attributable to the synchrotron peak, big blue bump emission, or extended emission. Blue data represent data we identified as being primarily attributable to inverse Compton emission. The green curve shows the fit to the red data; the solid parts show the portion within the range of the red data, while the dashed parts show the extension of the fit beyond the range of the red data. The black curve represents the same, but for the blue data. Frequencies are in Hertz. Fluxes are in erg cm\textsuperscript{-2} s\textsuperscript{-1}. Error bars are as they were presented by the original authors. Downward pointing arrows represent upper limits as they were presented by the original authors.}
        \label{planck_postage}
    \end{figure}

    \begin{figure}[h]
        \centering
        \includegraphics[width=\textwidth]{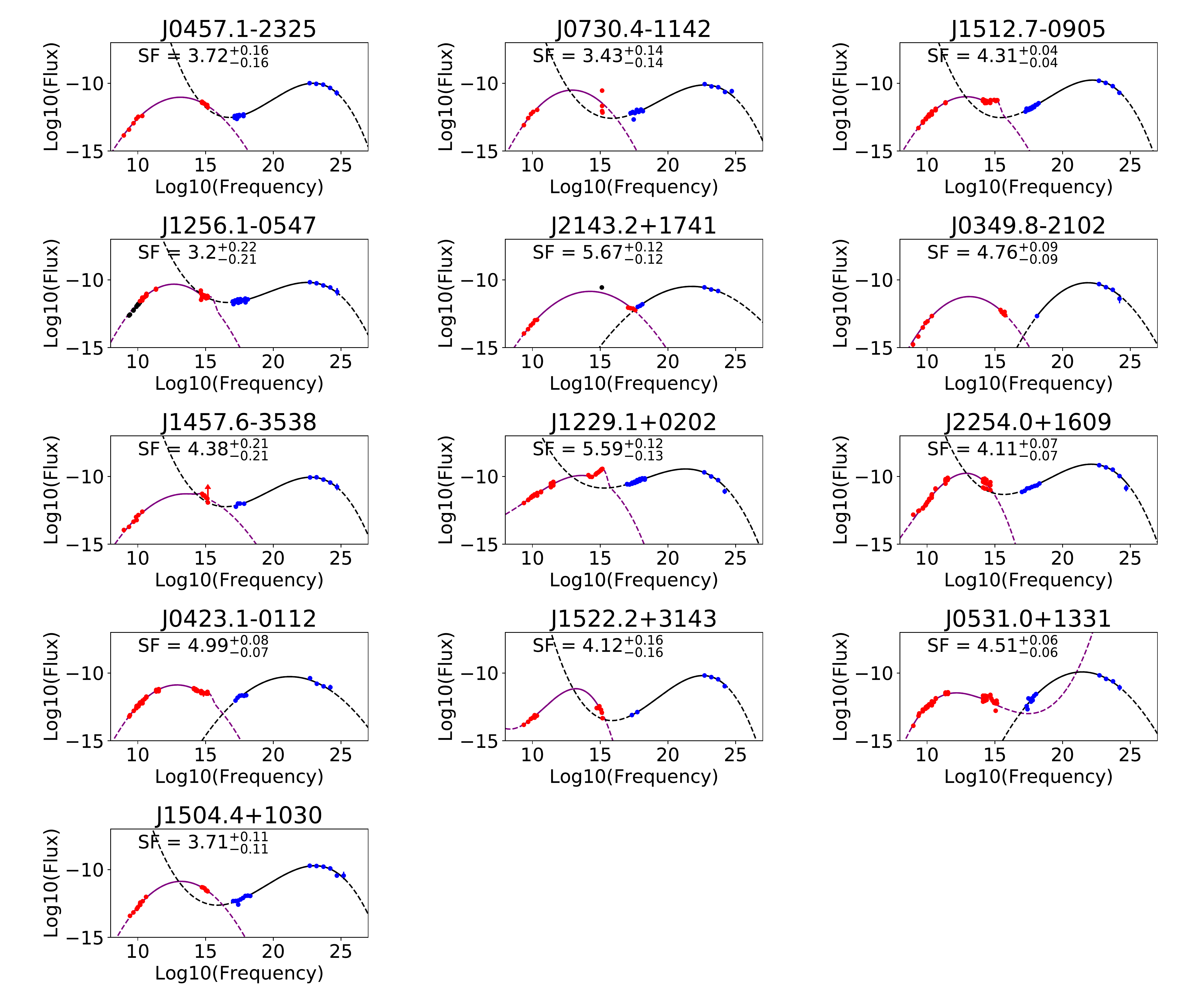}
        \caption{\textbf{LBAS Sample FSRQ SEDs.} The SEDs for the FSRQs in the LBAS Sample which passed all selection criteria. Red data points represent data we identified as being attributable to the synchrotron peak, big blue bump emission, or extended emission. Blue data represent data we identified as being primarily attributable to inverse Compton emission. The green curve shows the fit to the red data; the solid parts show the portion within the range of the red data, while the dashed parts show the extension of the fit beyond the range of the red data. The black curve represents the same, but for the blue data. Frequencies are in Hertz. Fluxes are in erg cm\textsuperscript{-2} s\textsuperscript{-1}. Error bars are as they were presented by the original authors. Downward pointing arrows represent upper limits as they were presented by the original authors.}
        \label{lbas_postage}
    \end{figure}

    \begin{figure}[h]
        \centering
        \includegraphics[width=\textwidth]{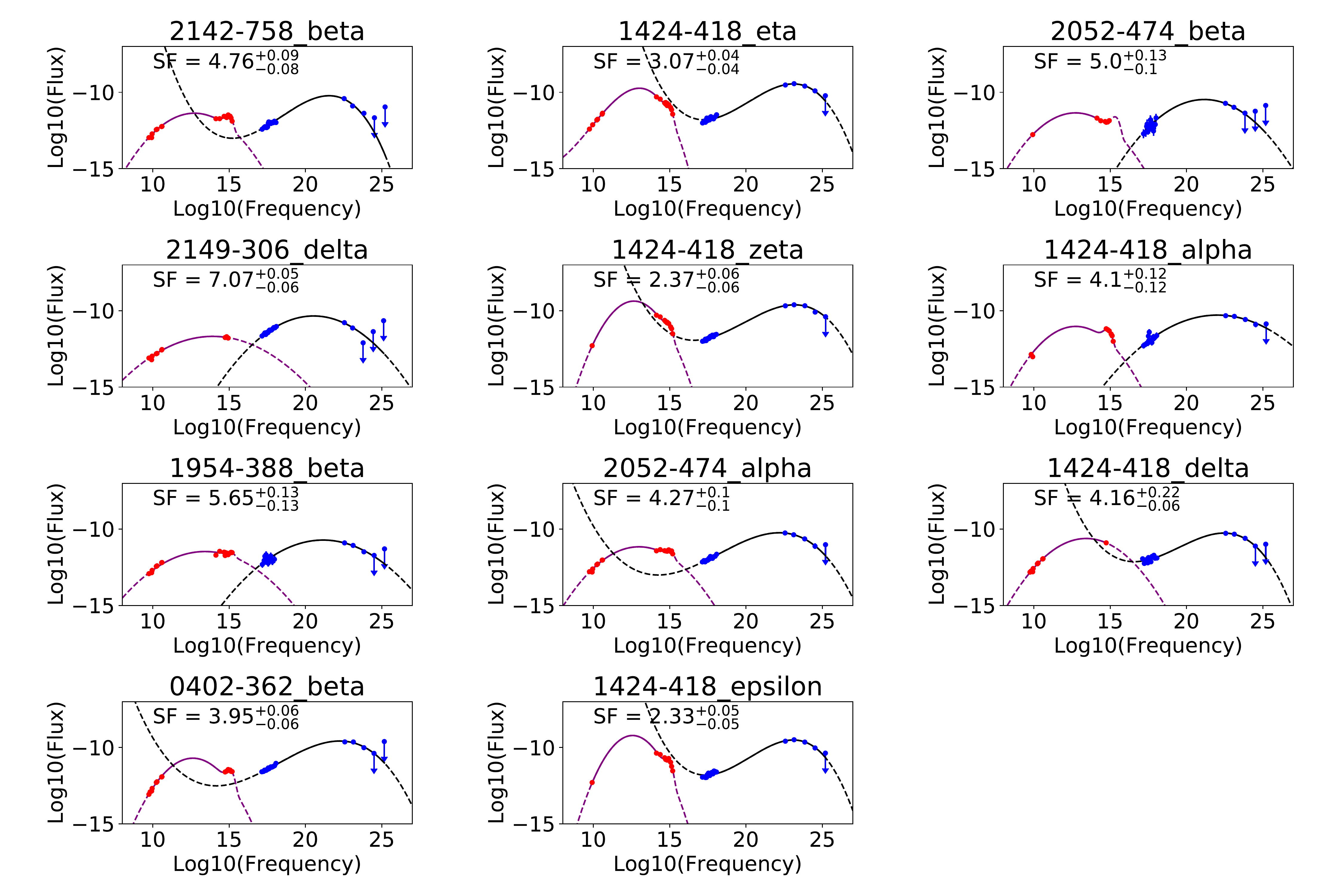}
        \caption{\textbf{DSSB Sample FSRQ SEDs.} The SEDs for the FSRQs in the DSSB Sample which passed all selection criteria. Red data points represent data we identified as being attributable to the synchrotron peak, big blue bump emission, or extended emission. Blue data represent data we identified as being primarily attributable to inverse Compton emission. The green curve shows the fit to the red data; the solid parts show the portion within the range of the red data, while the dashed parts show the extension of the fit beyond the range of the red data. The black curve represents the same, but for the blue data. Frequencies are in Hertz. Fluxes are in erg cm\textsuperscript{-2} s\textsuperscript{-1}. Error bars are as they were presented by the original authors. Upward/downward pointing arrows represent lower/upper limits as they were presented by the original authors.}
        \label{dssb_postage}
    \end{figure}

    \begin{figure}[h]
        \centering
        \includegraphics[width=\textwidth]{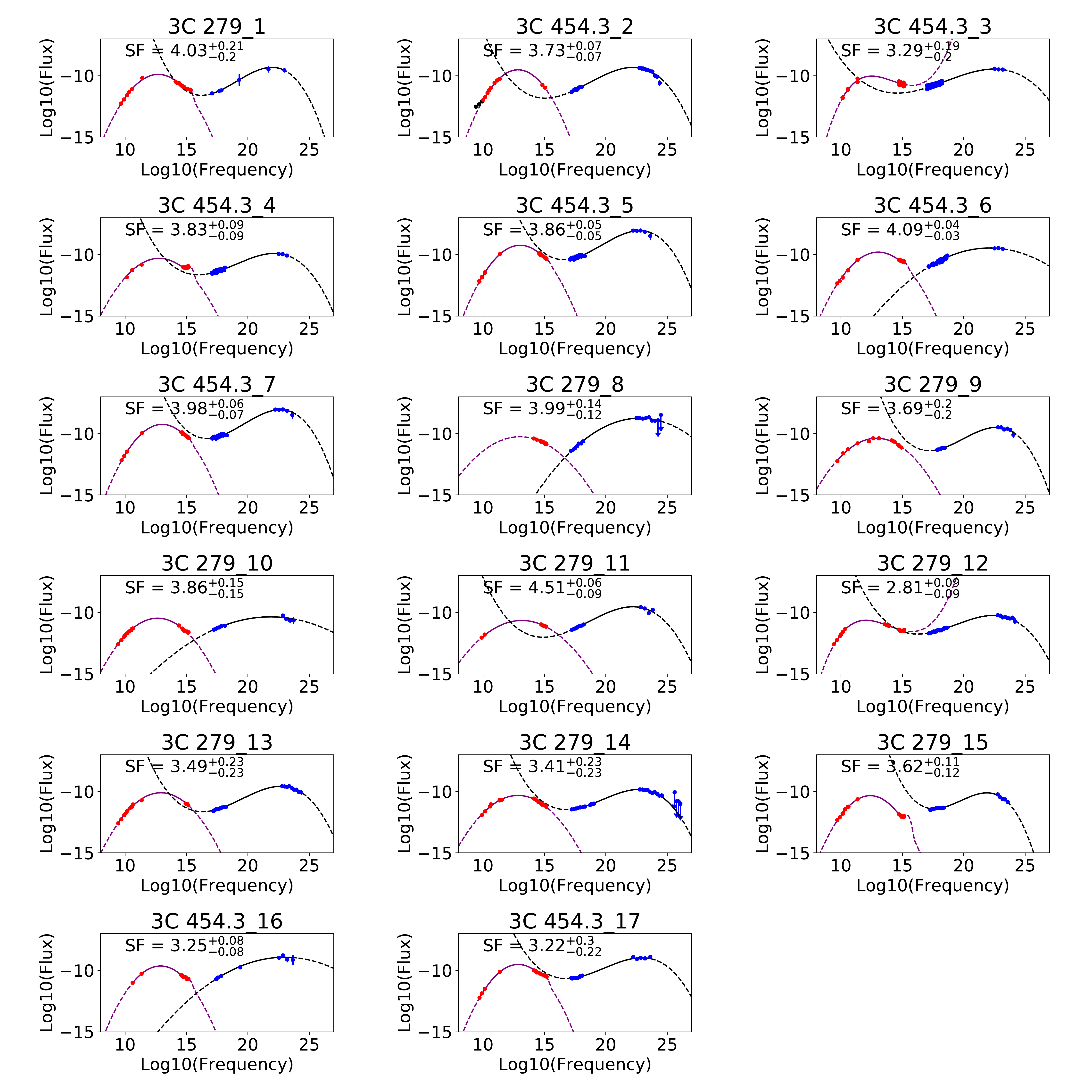}
        \caption{\textbf{LSS Sample FSRQ SEDs.} The SEDs for the FSRQs in the LSS Sample which passed all selection criteria. Red data points represent data we identified as being attributable to the synchrotron peak, big blue bump emission, or extended emission. Blue data represent data we identified as being primarily attributable to inverse Compton emission. The green curve shows the fit to the red data; the solid parts show the portion within the range of the red data, while the dashed parts show the extension of the fit beyond the range of the red data. The black curve represents the same, but for the blue data. Frequencies are in Hertz. Fluxes are in erg cm\textsuperscript{-2} s\textsuperscript{-1}. Error bars are as they were presented by the original authors. Naming is contracted to include only the source name and final trailing number as given in Supplementary Table 6. Downward pointing arrows represent upper limits as they were presented by the original authors.}
        \label{lss_postage}
    \end{figure}

\subsection{Impact of Variability in SED Samples}

The LBAS has a $\gamma $-ray integration time of $88$ days for all SEDs. A companion paper \cite{AbdoAPJ2010A} investigated the weekly lightcurves of the LBAS sources (note that $2$ of the sources in our paper are missing in their data) in a $47$ week period starting at the start of the LBAS period. We have calculated the normalized excess variance ($F_{var}=\sqrt{(S^{2}-\langle \sigma _{F}^{2} \rangle )/\langle F \rangle }$) for the $13$ weeks overlapping with the LBAS period, the normalized excess variance for the total $47$ weeks, and compared these two quantities by calculating the fractional difference between them. This is detailed in the table below. Of note are (1) the normalized excess variance for all sources is below $1$, with a maximum of about $0.8$, and a median of about $0.4$ (2) the fractional difference is only negative (i.e., the LBAS period has a larger excess variance than the total $47$ weeks) for one source, and the median of the fractional difference is about $0.3$, implying that the excess variance is generally a sizeable fraction smaller during the LBAS period than over the entire $47$ week period (i.e., there is less variability in the integration time of the LBAS than these sources experience on average).

 The DSSB SEDs were created via a Bayesian block method based on Fermi-LAT lightcurves to select periods of relatively steady flux.

 The other samples were integrated over similarly short lengths of time, and thus it is likely that variability is similarly small and/or averaged out.

\begin{table}
\centering
\caption{LBAS Normalized Excess Variances}
\begin{tabular}{|c|c|c|c|}
\hline
0FGL Name & LBAS Time EV & Total EV & Fractional Difference of EV \\
\hline
\hline
J0457.1-2325 & 0.23 & 0.48 & 0.53 \\
\hline
J1512.7-0905 & 0.80 & 0.95 & 0.15 \\
\hline
J1256.1-0547 & 0.15 & 0.78 & 0.81 \\
\hline
J2143.2+1741 & 0.37 & 0.51 & 0.27 \\
\hline
J0349.8-2102 & 0.27 & 0.71 & 0.61 \\
\hline
J1457.6-3538 & 0.58 & 0.60 & 0.04 \\
\hline
J1229.1+0202 & 0.51 & 0.61 & 0.17 \\
\hline
J2254.0+1609 & 0.35 & 0.92 & 0.62 \\
\hline
J1522.2+3143 & 0.26 & 0.30 & 0.16 \\
\hline
J0531.0+1331 & 0.43 & 0.81 & 0.47 \\
\hline
J1504.4+1030 & 0.63 & 0.40 & -0.57 \\
\hline
\end{tabular}
\end{table}

\subsection{Comparison with Seed Factor of Weak Extragalactic Jets}\label{SI:weak_jets}
    To test that the seed factor distribution for powerful extragalactic jets is different from sources in which there is no significant external photon field for external Compton scattering we calculate the seed factor for weak extragalactic jets.

    Beamed weak extragalactic jets generally correspond to BL Lacertae objects (BL Lacs, or BLLs). We have calculated the seed factor for sources in the LBAS Sample, Giommi Sample, and the DSSB Sample which are classified (in their respective papers) as BL Lacertae objects (see Supplementary Fig. \ref{bll_histogram}, and compare with Fig. 1). The number of BL Lac SEDs which we could reliably fit is $21$. This is about a third the number of FSRQ SEDs which we were able to reliably fit ($62$ SEDs for the FSRQs).

    The Kolmogorov-Smirnov test is not very sensitive to effects in the tails of a distribution. Global effects and the small sample size of the BL Lac sample reduce the reliability of the Kolmogorov-Smirnov test. We therefore applied the Anderson-Darling test and the Shapiro-Wilk test to large ($10^{7}$ variates) bootstrapped samples of the BL Lac and FSRQ seed factor distributions. Using a rejection significance threshold of $2\sigma $, the BL Lac distribution is significantly rejected as normal by both of these tests ($>2.58\sigma $ (p-value less than $0.01$) using the Anderson-Darling test, and $3.55\sigma $ \textbf{(p-value $0.00$)} using the Shapiro-Wilk test). The FSRQ distribution is not significantly rejected as normal by either of these tests ($1.64-1.96\sigma $ ($0.05 < $ p-value $< 0.1$) using the Anderson-Darling test, and $1.99\sigma $ (p-value $0.05$) using the Shapiro-Wilk test). These results imply that the BL Lacs and FSRQs have distinct seed factor distributions, and thus that the BL Lacs and FSRQs likely emit gamma-rays through different processes.

    A complicating factor is that BL Lacertae populations are known to be contaminated with ``fake BL Lacs'' \citep{GhiselliniMNRAS2011}, which are FSRQs which have been classified as BL Lacs due to their continuum emission outshining their broad-line emission. However, any contaminating FSRQs in our BL Lac sample are expected to have seed factors which behave the same as for the FSRQ sample (which is not a contaminated sample). Therefore, any differences between the FSRQ and BL Lac seed factor distributions are attributable to the real BL Lacs in the BL Lac sample.

    \begin{figure}
        \centering
        \includegraphics[width=\textwidth]{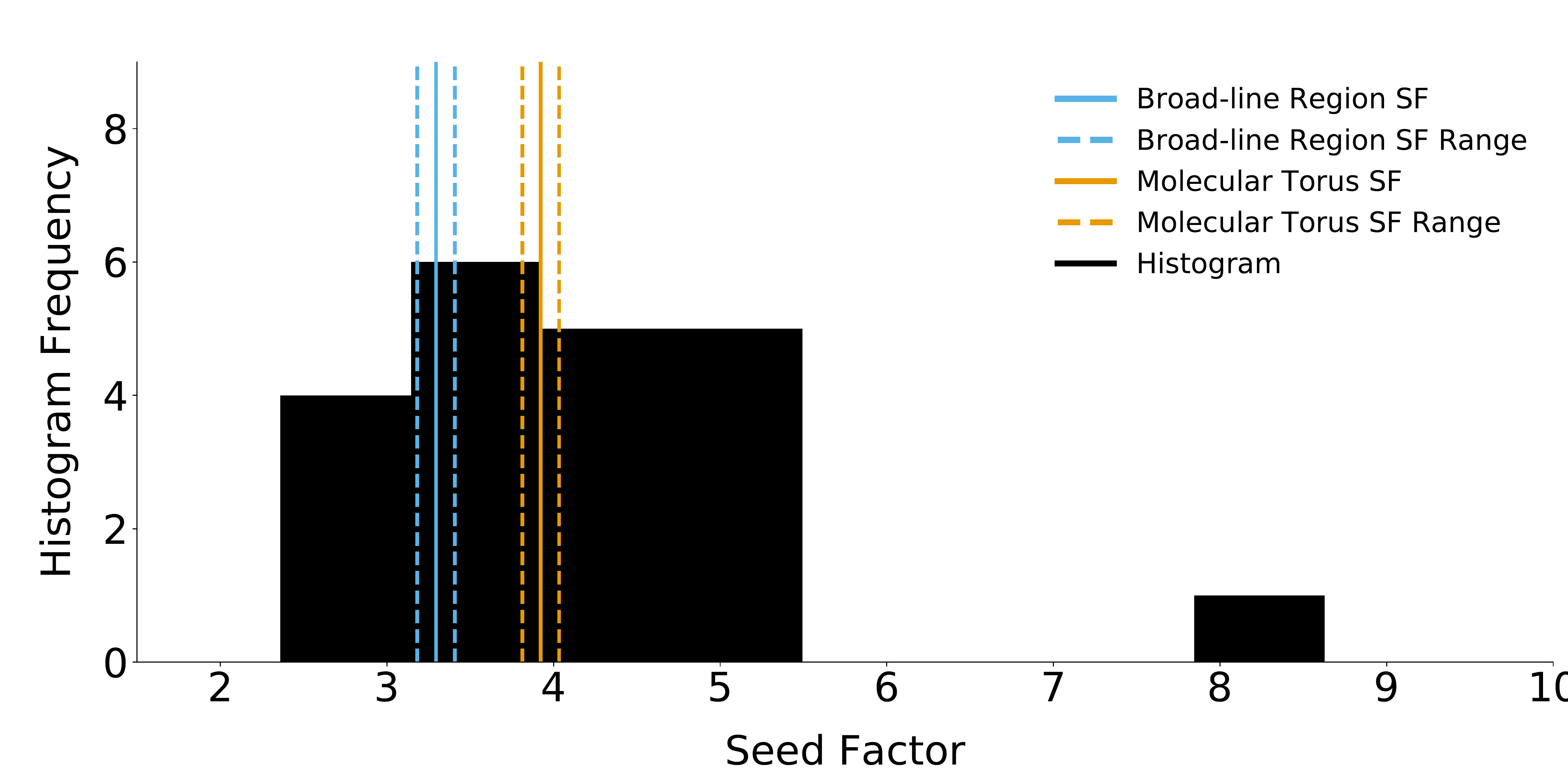}
        \caption{\textbf{Histogram of BL Lac seed factors.} A histogram showing the distribution of seed factors for our sample of BL Lacertae objects. We do not include a kernel density estimate here due to the small size of the BL Lac sample. For reference, the expected seed factor value and $1\sigma $ confidence interval for the broad-line region (blue) and molecular torus (orange) for FSRQs are shown by solid and dashed lines, respectively. See Sections 1.2 and 1.3 of the Supplementary Information for information on calculation of the plotted confidence intervals. The distribution of seed factors  appears flat for BL Lacs as compared to for FSRQs.  Compared with Fig. 1, it is clear that the FSRQ seed factor distribution has a sharper peak than the BL Lac seed factor distribution.}
        \label{bll_histogram}
    \end{figure}

\FloatBarrier
\clearpage
\subsection{Seed Factor Tables}
\nopagebreak
\begin{table}[h!]
\centering
\caption{DSSB FSRQ Table\label{dssb_fsrq_table}}
\rotatebox{0}{\begin{tabular}{|c|c|c|c|c|c|c|c|c|c|c|c|c|c|c|c|c|}
\hline
Source & SED Name & z & SF & $\text{Log}10(\nu_{s})$ & $\text{Log}10(\nu_{s}F_{\nu_{s}})$ & $\text{Log}10(\nu_{IC})$ & $\text{Log}10(\nu_{IC}F_{\nu_{IC}})$ \\
\hline
\hline
2142-758 & 2142-758\_beta & 1.14 & $4.76_{-0.08}^{+0.09}$ & $12.73_{-0.01}^{+0.01}$ & $-11.37_{-0.01}^{+0.01}$ & $21.55_{-0.08}^{+0.08}$ & $-10.22_{-0.05}^{+0.05}$ \\
\hline
1424-418 & 1424-418\_eta & 1.52 & $3.07_{-0.04}^{+0.04}$ & $13.02_{-0.02}^{+0.02}$ & $-9.73_{-0.02}^{+0.02}$ & $23.10_{-0.04}^{+0.04}$ & $-9.45_{-0.01}^{+0.01}$ \\
\hline
2052-474 & 2052-474\_beta & 1.49 & $5.00_{-0.10}^{+0.13}$ & $12.73_{-0.04}^{+0.04}$ & $-11.35_{-0.03}^{+0.03}$ & $21.17_{-0.12}^{+0.09}$ & $-10.47_{-0.07}^{+0.06}$ \\
\hline
2149-306 & 2149-306\_delta & 2.35 & $7.07_{-0.06}^{+0.05}$ & $13.93_{-0.02}^{+0.02}$ & $-11.68_{-0.01}^{+0.01}$ & $20.54_{-0.05}^{+0.05}$ & $-10.34_{-0.03}^{+0.03}$ \\
\hline
1424-418 & 1424-418\_zeta & 1.52 & $2.37_{-0.06}^{+0.06}$ & $12.66_{-0.01}^{+0.01}$ & $-9.37_{-0.02}^{+0.02}$ & $23.18_{-0.05}^{+0.05}$ & $-9.61_{-0.01}^{+0.01}$ \\
\hline
1424-418 & 1424-418\_alpha & 1.52 & $4.10_{-0.12}^{+0.12}$ & $12.74_{-0.01}^{+0.01}$ & $-11.03_{-0.01}^{+0.01}$ & $22.02_{-0.12}^{+0.12}$ & $-10.29_{-0.04}^{+0.04}$ \\
\hline
1954-388 & 1954-388\_beta & 0.63 & $5.65_{-0.13}^{+0.13}$ & $13.42_{-0.04}^{+0.04}$ & $-11.47_{-0.02}^{+0.02}$ & $21.16_{-0.12}^{+0.11}$ & $-10.72_{-0.06}^{+0.06}$ \\
\hline
2052-474 & 2052-474\_alpha & 1.49 & $4.27_{-0.10}^{+0.10}$ & $12.99_{-0.03}^{+0.03}$ & $-11.16_{-0.02}^{+0.02}$ & $22.19_{-0.10}^{+0.09}$ & $-10.24_{-0.08}^{+0.07}$ \\
\hline
1424-418 & 1424-418\_delta & 1.52 & $4.16_{-0.06}^{+0.22}$ & $13.42_{-0.00}^{+0.00}$ & $-10.61_{-0.00}^{+0.00}$ & $22.45_{-0.22}^{+0.05}$ & $-10.26_{-0.07}^{+0.01}$ \\
\hline
0402-362 & 0402-362\_beta & 1.42 & $3.95_{-0.06}^{+0.06}$ & $12.62_{-0.01}^{+0.01}$ & $-10.70_{-0.02}^{+0.02}$ & $22.24_{-0.06}^{+0.06}$ & $-9.57_{-0.04}^{+0.04}$ \\
\hline
1424-418 & 1424-418\_epsilon & 1.52 & $2.33_{-0.05}^{+0.05}$ & $12.57_{-0.02}^{+0.02}$ & $-9.21_{-0.04}^{+0.04}$ & $23.10_{-0.04}^{+0.04}$ & $-9.51_{-0.01}^{+0.01}$ \\
\hline
\end{tabular}}
\end{table}

\FloatBarrier
\begin{table}[h!]
\centering
\caption{LBAS FSRQ Table\label{lbas_fsrq_table}}
\rotatebox{0}{\begin{tabular}{|c|c|c|c|c|c|c|c|c|c|c|c|c|c|c|c|c|}
\hline
Source & SED Name & z & SF & $\text{Log}10(\nu_{s})$ & $\text{Log}10(\nu_{s}F_{\nu_{s}})$ & $\text{Log}10(\nu_{IC})$ & $\text{Log}10(\nu_{IC}F_{\nu_{IC}})$ \\
\hline
\hline
J0457.1-2325 & J0457.1-2325 & 1.00 & $3.72_{-0.16}^{+0.16}$ & $13.14_{-0.02}^{+0.02}$ & $-11.03_{-0.02}^{+0.02}$ & $22.95_{-0.15}^{+0.15}$ & $-10.00_{-0.04}^{+0.04}$ \\
\hline
J0730.4-1142 & J0730.4-1142 & 1.59 & $3.43_{-0.14}^{+0.14}$ & $12.94_{-0.02}^{+0.02}$ & $-10.50_{-0.02}^{+0.02}$ & $22.71_{-0.14}^{+0.14}$ & $-10.12_{-0.06}^{+0.05}$ \\
\hline
J1512.7-0905 & J1512.7-0905 & 0.36 & $4.31_{-0.04}^{+0.04}$ & $12.89_{-0.01}^{+0.01}$ & $-10.99_{-0.01}^{+0.01}$ & $22.20_{-0.03}^{+0.03}$ & $-9.76_{-0.02}^{+0.02}$ \\
\hline
J1256.1-0547 & J1256.1-0547 & 0.54 & $3.20_{-0.21}^{+0.22}$ & $12.67_{-0.02}^{+0.02}$ & $-10.32_{-0.02}^{+0.02}$ & $22.53_{-0.21}^{+0.20}$ & $-10.19_{-0.07}^{+0.08}$ \\
\hline
J2143.2+1741 & J2143.2+1741 & 0.21 & $5.67_{-0.12}^{+0.12}$ & $14.25_{-0.02}^{+0.02}$ & $-10.85_{-0.02}^{+0.02}$ & $21.77_{-0.12}^{+0.11}$ & $-10.48_{-0.05}^{+0.05}$ \\
\hline
J0349.8-2102 & J0349.8-2102 & 2.94 & $4.76_{-0.09}^{+0.09}$ & $13.12_{-0.02}^{+0.02}$ & $-11.23_{-0.02}^{+0.02}$ & $21.87_{-0.09}^{+0.09}$ & $-10.21_{-0.05}^{+0.05}$ \\
\hline
J1457.6-3538 & J1457.6-3538 & 1.42 & $4.38_{-0.21}^{+0.21}$ & $13.52_{-0.03}^{+0.03}$ & $-11.28_{-0.03}^{+0.02}$ & $22.76_{-0.20}^{+0.21}$ & $-10.06_{-0.08}^{+0.08}$ \\
\hline
J1229.1+0202 & J1229.1+0202 & 0.16 & $5.59_{-0.13}^{+0.12}$ & $13.63_{-0.10}^{+0.09}$ & $-9.93_{-0.04}^{+0.04}$ & $21.29_{-0.08}^{+0.09}$ & $-9.45_{-0.04}^{+0.04}$ \\
\hline
J2254.0+1609 & J2254.0+1609 & 0.86 & $4.11_{-0.07}^{+0.07}$ & $12.88_{-0.02}^{+0.02}$ & $-9.77_{-0.02}^{+0.02}$ & $22.11_{-0.07}^{+0.10}$ & $-9.09_{-0.05}^{+0.04}$ \\
\hline
J0423.1-0112 & J0423.1-0112 & 0.92 & $4.99_{-0.07}^{+0.08}$ & $12.89_{-0.01}^{+0.01}$ & $-10.88_{-0.01}^{+0.01}$ & $21.22_{-0.07}^{+0.07}$ & $-10.26_{-0.04}^{+0.04}$ \\
\hline
J1522.2+3143 & J1522.2+3143 & 1.49 & $4.12_{-0.16}^{+0.16}$ & $13.21_{-0.04}^{+0.04}$ & $-11.15_{-0.06}^{+0.07}$ & $22.58_{-0.14}^{+0.15}$ & $-10.19_{-0.06}^{+0.06}$ \\
\hline
J0531.0+1331 & J0531.0+1331 & 2.07 & $4.51_{-0.06}^{+0.06}$ & $12.17_{-0.03}^{+0.03}$ & $-11.46_{-0.02}^{+0.02}$ & $21.43_{-0.04}^{+0.05}$ & $-9.92_{-0.04}^{+0.04}$ \\
\hline
J1504.4+1030 & J1504.4+1030 & 1.84 & $3.71_{-0.11}^{+0.11}$ & $13.19_{-0.01}^{+0.01}$ & $-9.71_{-0.04}^{+0.04}$ & $23.06_{-0.11}^{+0.11}$ & $-9.71_{-0.04}^{+0.04}$ \\
\hline
\end{tabular}}
\end{table}

\FloatBarrier
\begin{table}[h!]
\centering
\caption{Giommi FSRQ Table\label{giommi_fsrq_table}}
\rotatebox{0}{\begin{tabular}{|c|c|c|c|c|c|c|c|c|c|c|c|c|c|c|c|c|}
\hline
Source & SED Name & z & SF & $\text{Log}10(\nu_{s})$ & $\text{Log}10(\nu_{s}F_{\nu_{s}})$ & $\text{Log}10(\nu_{IC})$ & $\text{Log}10(\nu_{IC}F_{\nu_{IC}})$ \\
\hline
\hline
J0457-2324 & J0457-2324 & 1.00 & $3.31_{-0.13}^{+0.13}$ & $12.97_{-0.04}^{+0.04}$ & $-11.35_{-0.02}^{+0.02}$ & $23.14_{-0.12}^{+0.12}$ & $-10.41_{-0.07}^{+0.07}$ \\
\hline
J1256-0547 & J1256-0547 & 0.54 & $2.58_{-0.10}^{+0.10}$ & $12.45_{-0.02}^{+0.02}$ & $-10.48_{-0.02}^{+0.02}$ & $22.97_{-0.09}^{+0.09}$ & $-10.29_{-0.04}^{+0.04}$ \\
\hline
J0920+4441 & J0920+4441 & 2.19 & $3.11_{-0.13}^{+0.13}$ & $11.89_{-0.03}^{+0.03}$ & $-11.35_{-0.04}^{+0.04}$ & $22.21_{-0.12}^{+0.12}$ & $-10.53_{-0.07}^{+0.07}$ \\
\hline
J0136+4751 & J0136+4751 & 0.86 & $7.31_{-0.11}^{+0.12}$ & $12.85_{-0.02}^{+0.03}$ & $-10.90_{-0.03}^{+0.01}$ & $18.01_{-0.11}^{+0.10}$ & $-11.99_{-0.06}^{+0.05}$ \\
\hline
J1642+3948 & J1642+3948 & 0.59 & $2.40_{-0.09}^{+0.09}$ & $11.73_{-0.04}^{+0.03}$ & $-11.31_{-0.02}^{+0.02}$ & $22.60_{-0.08}^{+0.08}$ & $-10.78_{-0.03}^{+0.03}$ \\
\hline
J1924-2914 & J1924-2914 & 0.35 & $1.92_{-0.13}^{+0.12}$ & $12.48_{-0.05}^{+0.05}$ & $-10.76_{-0.03}^{+0.03}$ & $23.59_{-0.11}^{+0.11}$ & $-10.71_{-0.03}^{+0.03}$ \\
\hline
J1911-2006 & J1911-2006 & 1.12 & $3.97_{-0.14}^{+0.13}$ & $12.54_{-0.04}^{+0.04}$ & $-11.20_{-0.03}^{+0.03}$ & $21.76_{-0.12}^{+0.13}$ & $-10.84_{-0.09}^{+0.08}$ \\
\hline
J0010+1058 & J0010+1058 & 0.09 & $6.54_{-0.14}^{+0.14}$ & $14.36_{-0.07}^{+0.07}$ & $-10.73_{-0.02}^{+0.02}$ & $20.84_{-0.11}^{+0.11}$ & $-10.70_{-0.04}^{+0.04}$ \\
\hline
J0530+1331 & J0530+1331 & 2.07 & $5.62_{-0.05}^{+0.02}$ & $13.85_{-0.03}^{+0.00}$ & $-11.26_{-0.02}^{+0.00}$ & $21.63_{-0.07}^{+0.00}$ & $-10.48_{-0.06}^{+0.00}$ \\
\hline
J1130-1449 & J1130-1449 & 1.18 & $5.26_{-0.11}^{+0.11}$ & $12.69_{-0.04}^{+0.04}$ & $-11.14_{-0.03}^{+0.03}$ & $20.47_{-0.10}^{+0.10}$ & $-11.09_{-0.04}^{+0.04}$ \\
\hline
J2232+1143 & J2232+1143 & 1.04 & $2.95_{-0.13}^{+0.15}$ & $11.46_{-0.05}^{+0.04}$ & $-11.59_{-0.03}^{+0.03}$ & $21.82_{-0.14}^{+0.11}$ & $-10.98_{-0.06}^{+0.05}$ \\
\hline
J2229-0832 & J2229-0832 & 1.56 & $4.62_{-0.08}^{+0.07}$ & $12.71_{-0.04}^{+0.04}$ & $-11.34_{-0.04}^{+0.04}$ & $21.33_{-0.06}^{+0.06}$ & $-10.88_{-0.02}^{+0.02}$ \\
\hline
J1923-2104 & J1923-2104 & 0.87 & $3.65_{-0.08}^{+0.16}$ & $13.46_{-0.06}^{+0.06}$ & $-10.94_{-0.03}^{+0.03}$ & $22.99_{-0.12}^{+0.07}$ & $-10.59_{-0.04}^{+0.07}$ \\
\hline
J0957+5522 & J0957+5522 & 0.90 & $5.60_{-0.25}^{+0.24}$ & $13.97_{-0.20}^{+0.26}$ & $-11.88_{-0.04}^{+0.04}$ & $21.85_{-0.14}^{+0.13}$ & $-10.95_{-0.09}^{+0.09}$ \\
\hline
J0237+2848 & J0237+2848 & 1.21 & $4.51_{-0.05}^{+0.20}$ & $12.78_{-0.05}^{+0.05}$ & $-11.44_{-0.03}^{+0.03}$ & $21.75_{-0.12}^{+0.00}$ & $-10.50_{-0.06}^{+0.00}$ \\
\hline
J1222+0413 & J1222+0413 & 0.97 & $4.16_{-0.08}^{+0.08}$ & $11.83_{-0.05}^{+0.05}$ & $-11.74_{-0.04}^{+0.04}$ & $21.33_{-0.06}^{+0.06}$ & $-10.45_{-0.05}^{+0.05}$ \\
\hline
J2203+3145 & J2203+3145 & 0.29 & $5.37_{-0.09}^{+0.09}$ & $12.07_{-0.03}^{+0.03}$ & $-11.31_{-0.03}^{+0.03}$ & $19.82_{-0.08}^{+0.07}$ & $-11.07_{-0.03}^{+0.03}$ \\
\hline
J2148+0657 & J2148+0657 & 1.00 & $4.61_{-0.10}^{+0.10}$ & $12.89_{-0.03}^{+0.03}$ & $-11.02_{-0.02}^{+0.02}$ & $21.25_{-0.09}^{+0.09}$ & $-11.10_{-0.03}^{+0.03}$ \\
\hline
J2207-5346 & J2207-5346 & 1.22 & $4.48_{-0.11}^{+0.10}$ & $11.83_{-0.05}^{+0.05}$ & $-12.01_{-0.04}^{+0.04}$ & $20.99_{-0.08}^{+0.08}$ & $-10.74_{-0.08}^{+0.08}$ \\
\hline
J1504+1029 & J1504+1029 & 1.84 & $4.09_{-0.26}^{+0.23}$ & $12.94_{-0.02}^{+0.09}$ & $-11.88_{-0.04}^{+0.01}$ & $22.53_{-0.23}^{+0.24}$ & $-10.55_{-0.08}^{+0.09}$ \\
\hline
\end{tabular}}
\end{table}

\FloatBarrier
\begin{table}[h!]
\centering
\caption{Planck FSRQ Table\label{planck_fsrq_table}}
\rotatebox{0}{\begin{tabular}{|c|c|c|c|c|c|c|c|c|c|c|c|c|c|c|c|c|}
\hline
Source & SED Name & z & SF & $\text{Log}10(\nu_{s})$ & $\text{Log}10(\nu_{s}F_{\nu_{s}})$ & $\text{Log}10(\nu_{IC})$ & $\text{Log}10(\nu_{IC}F_{\nu_{IC}})$ \\
\hline
\hline
J1224+2122 & J1224+2122 & 0.43 & $3.17_{-0.26}^{+0.27}$ & $12.01_{-0.02}^{+0.02}$ & $-11.59_{-0.03}^{+0.03}$ & $22.69_{-0.26}^{+0.27}$ & $-9.92_{-0.09}^{+0.10}$ \\
\hline
\end{tabular}}
\end{table}

\FloatBarrier
\begin{table}[h!]
\centering
\caption{LSS FSRQ Table\label{lss_fsrq_table}}
\rotatebox{0}{\begin{tabular}{|c|c|c|c|c|c|c|c|c|c|c|c|c|c|c|c|c|}
\hline
Source & SED Name & z & SF & $\text{Log}10(\nu_{s})$ & $\text{Log}10(\nu_{s}F_{\nu_{s}})$ & $\text{Log}10(\nu_{IC})$ & $\text{Log}10(\nu_{IC}F_{\nu_{IC}})$ \\
\hline
\hline
3C 279 & 3C\_279\_January\_1 \citep{WehrleAPJ1998} & 0.54 & $4.03_{-0.20}^{+0.21}$ & $12.70_{-0.01}^{+0.01}$ & $-9.89_{-0.01}^{+0.01}$ & $21.96_{-0.19}^{+0.22}$ & $-9.32_{-0.09}^{+0.10}$ \\
\hline
3C 454.3 & 3C\_454.3\_2 \citep{AbdoAPJ2009} & 0.86 & $3.73_{-0.07}^{+0.07}$ & $12.85_{-0.01}^{+0.02}$ & $-9.52_{-0.02}^{+0.02}$ & $22.23_{-0.06}^{+0.06}$ & $-9.32_{-0.04}^{+0.04}$ \\
\hline
3C 454.3 & 3C\_454.3\_54673\_54693\_3 \citep{VercelloneAPJ2010} & 0.86 & $3.29_{-0.20}^{+0.19}$ & $12.50_{-0.03}^{+0.03}$ & $-10.03_{-0.01}^{+0.02}$ & $22.51_{-0.19}^{+0.19}$ & $-9.46_{-0.08}^{+0.08}$ \\
\hline
3C 454.3 & 3C\_454.3\_54800\_54845\_4 \citep{VercelloneAPJ2010} & 0.86 & $3.83_{-0.09}^{+0.09}$ & $12.77_{-0.02}^{+0.02}$ & $-10.29_{-0.02}^{+0.01}$ & $22.15_{-0.09}^{+0.09}$ & $-9.90_{-0.03}^{+0.03}$ \\
\hline
3C 454.3 & 3C\_454.3\_flare\_5 \citep{VercelloneAPJ2011} & 0.86 & $3.86_{-0.05}^{+0.05}$ & $13.02_{-0.00}^{+0.00}$ & $-9.23_{-0.01}^{+0.01}$ & $22.76_{-0.05}^{+0.04}$ & $-8.04_{-0.02}^{+0.02}$ \\
\hline
3C 454.3 & 3C\_454.3\_54617\_54618\_6 \citep{VercelloneAPJ2010} & 0.86 & $4.09_{-0.03}^{+0.04}$ & $13.04_{-0.02}^{+0.02}$ & $-9.79_{-0.02}^{+0.02}$ & $22.12_{-0.02}^{+0.02}$ & $-9.45_{-0.01}^{+0.01}$ \\
\hline
3C 454.3 & 3C\_454.3\_post-flare\_7 \citep{VercelloneAPJ2011} & 0.86 & $3.98_{-0.07}^{+0.06}$ & $13.02_{-0.01}^{+0.01}$ & $-9.24_{-0.01}^{+0.01}$ & $22.64_{-0.06}^{+0.06}$ & $-8.05_{-0.02}^{+0.02}$ \\
\hline
3C 279 & 3C\_279\_Period\_D\_8 \citep{HayashidaAPJ2015} & 0.54 & $3.99_{-0.12}^{+0.14}$ & $13.01_{-0.05}^{+0.06}$ & $-10.25_{-0.02}^{+0.02}$ & $22.79_{-0.06}^{+0.04}$ & $-8.74_{-0.04}^{+0.03}$ \\
\hline
3C 279 & 3C\_279\_Period\_D\_9 \citep{HayashidaAPJ2012} & 0.54 & $3.69_{-0.20}^{+0.20}$ & $12.91_{-0.03}^{+0.05}$ & $-10.37_{-0.02}^{+0.04}$ & $22.68_{-0.13}^{+0.17}$ & $-9.48_{-0.06}^{+0.08}$ \\
\hline
3C 279 & 3C\_279\_Period\_F\_10 \citep{HayashidaAPJ2012} & 0.54 & $3.86_{-0.15}^{+0.15}$ & $12.65_{-0.01}^{+0.02}$ & $-10.45_{-0.02}^{+0.01}$ & $21.85_{-0.08}^{+0.10}$ & $-10.34_{-0.04}^{+0.03}$ \\
\hline
3C 279 & 3C\_279\_Period\_G\_11 \citep{HayashidaAPJ2012} & 0.54 & $4.51_{-0.09}^{+0.06}$ & $13.15_{-0.06}^{+0.04}$ & $-10.64_{-0.02}^{+0.03}$ & $22.21_{-0.05}^{+0.05}$ & $-9.52_{-0.02}^{+0.02}$ \\
\hline
3C 279 & 3C\_279\_Period\_A\_12 \citep{HayashidaAPJ2012} & 0.54 & $2.81_{-0.09}^{+0.09}$ & $12.04_{-0.05}^{+0.06}$ & $-10.62_{-0.06}^{+0.05}$ & $22.43_{-0.07}^{+0.07}$ & $-10.23_{-0.02}^{+0.02}$ \\
\hline
3C 279 & 3C\_279\_Period\_B\_13 \citep{HayashidaAPJ2012} & 0.54 & $3.49_{-0.23}^{+0.23}$ & $12.91_{-0.03}^{+0.03}$ & $-10.09_{-0.03}^{+0.02}$ & $22.69_{-0.17}^{+0.17}$ & $-9.57_{-0.10}^{+0.10}$ \\
\hline
3C 279 & 3C\_279\_Period\_C\_14 \citep{HayashidaAPJ2012} & 0.54 & $3.41_{-0.23}^{+0.23}$ & $12.85_{-0.03}^{+0.03}$ & $-10.31_{-0.02}^{+0.02}$ & $22.70_{-0.18}^{+0.19}$ & $-9.83_{-0.06}^{+0.07}$ \\
\hline
3C 279 & 3C\_279\_Period\_H\_15 \citep{HayashidaAPJ2012} & 0.54 & $3.62_{-0.12}^{+0.11}$ & $12.38_{-0.02}^{+0.02}$ & $-10.33_{-0.02}^{+0.03}$ & $21.88_{-0.11}^{+0.11}$ & $-10.10_{-0.08}^{+0.09}$ \\
\hline
3C 454.3 & 3C\_454.3\_pre-flare\_16 \citep{PaccianiAPJ2010} & 0.86 & $3.25_{-0.08}^{+0.08}$ & $12.89_{-0.03}^{+0.03}$ & $-9.63_{-0.03}^{+0.03}$ & $23.00_{-0.07}^{+0.07}$ & $-8.91_{-0.04}^{+0.04}$ \\
\hline
3C 454.3 & 3C\_454.3\_pre-flare\_17 \citep{VercelloneAPJ2011} & 0.86 & $3.22_{-0.22}^{+0.30}$ & $12.87_{-0.02}^{+0.02}$ & $-9.50_{-0.02}^{+0.03}$ & $22.93_{-0.31}^{+0.18}$ & $-8.96_{-0.08}^{+0.07}$ \\
\hline
\end{tabular}}
\end{table}

\FloatBarrier
\begin{table}[h!]
\centering
\caption{DSSB BL Lac Table\label{dssb_bll_table}}
\rotatebox{0}{\begin{tabular}{|c|c|c|c|c|c|c|c|c|c|c|c|c|c|c|c|c|}
\hline
Source & SED Name & z & SF & $\text{Log}10(\nu_{s})$ & $\text{Log}10(\nu_{s}F_{\nu_{s}})$ & $\text{Log}10(\nu_{IC})$ & $\text{Log}10(\nu_{IC}F_{\nu_{IC}})$ \\
\hline
\hline
0537-441 & 0537-441\_alpha & 0.89 & $3.93_{-0.06}^{+0.06}$ & $13.21_{-0.01}^{+0.01}$ & $-10.61_{-0.01}^{+0.01}$ & $22.48_{-0.06}^{+0.06}$ & $-10.22_{-0.03}^{+0.03}$ \\
\hline
0537-441 & 0537-441\_zeta & 0.89 & $4.15_{-0.16}^{+0.15}$ & $13.24_{-0.00}^{+0.00}$ & $-10.70_{-0.00}^{+0.00}$ & $22.24_{-0.16}^{+0.16}$ & $-10.42_{-0.05}^{+0.05}$ \\
\hline
0426-380 & 0426-380\_epsilon & 1.11 & $3.55_{-0.08}^{+0.08}$ & $13.61_{-0.02}^{+0.02}$ & $-10.68_{-0.02}^{+0.02}$ & $23.34_{-0.08}^{+0.08}$ & $-10.13_{-0.02}^{+0.02}$ \\
\hline
0537-441 & 0537-441\_delta & 0.89 & $2.36_{-0.04}^{+0.04}$ & $12.87_{-0.00}^{+0.00}$ & $-9.81_{-0.00}^{+0.00}$ & $23.50_{-0.04}^{+0.04}$ & $-9.84_{-0.02}^{+0.02}$ \\
\hline
0537-441 & 0537-441\_epsilon & 0.89 & $3.22_{-0.23}^{+0.23}$ & $13.44_{-0.00}^{+0.00}$ & $-10.46_{-0.00}^{+0.00}$ & $23.37_{-0.22}^{+0.23}$ & $-10.15_{-0.08}^{+0.08}$ \\
\hline
0426-380 & 0426-380\_gamma & 1.11 & $3.55_{-0.28}^{+0.28}$ & $13.47_{-0.02}^{+0.02}$ & $-11.30_{-0.01}^{+0.01}$ & $23.29_{-0.28}^{+0.29}$ & $-10.60_{-0.06}^{+0.06}$ \\
\hline
1057-797 & 1057-797\_alpha & 0.58 & $4.71_{-0.19}^{+0.19}$ & $13.08_{-0.00}^{+0.00}$ & $-11.47_{-0.00}^{+0.00}$ & $21.62_{-0.19}^{+0.19}$ & $-10.98_{-0.06}^{+0.06}$ \\
\hline
0447-439 & 0447-439\_gamma & 0.11 & $3.89_{-0.19}^{+0.20}$ & $15.02_{-0.05}^{+0.05}$ & $-10.26_{-0.01}^{+0.01}$ & $23.85_{-0.19}^{+0.16}$ & $-10.81_{-0.05}^{+0.05}$ \\
\hline
0537-441 & 0537-441\_gamma & 0.89 & $4.05_{-0.08}^{+0.08}$ & $13.36_{-0.02}^{+0.02}$ & $-10.33_{-0.01}^{+0.01}$ & $22.40_{-0.07}^{+0.07}$ & $-10.16_{-0.03}^{+0.03}$ \\
\hline
0537-441 & 0537-441\_beta & 0.89 & $4.83_{-0.07}^{+0.07}$ & $13.55_{-0.03}^{+0.03}$ & $-11.12_{-0.01}^{+0.01}$ & $22.13_{-0.06}^{+0.10}$ & $-10.34_{-0.03}^{+0.03}$ \\
\hline
\end{tabular}}
\end{table}

\FloatBarrier
\begin{table}[h!]
\centering
\caption{LBAS BL Lac Table\label{lbas_bll_table}}
\rotatebox{0}{\begin{tabular}{|c|c|c|c|c|c|c|c|c|c|c|c|c|c|c|c|c|}
\hline
Source & SED Name & z & SF & $\text{Log}10(\nu_{s})$ & $\text{Log}10(\nu_{s}F_{\nu_{s}})$ & $\text{Log}10(\nu_{IC})$ & $\text{Log}10(\nu_{IC}F_{\nu_{IC}})$ \\
\hline
\hline
J1751.5+0935 & J1751.5+0935 & 0.32 & $2.99_{-0.25}^{+0.25}$ & $12.87_{-0.01}^{+0.01}$ & $-10.22_{-0.02}^{+0.02}$ & $22.80_{-0.24}^{+0.21}$ & $-10.39_{-0.10}^{+0.07}$ \\
\hline
J0712.9+5034 & J0712.9+5034 & 0.50 & $3.10_{-0.18}^{+0.18}$ & $13.55_{-0.04}^{+0.04}$ & $-11.21_{-0.02}^{+0.02}$ & $23.58_{-0.17}^{+0.16}$ & $-10.98_{-0.07}^{+0.08}$ \\
\hline
J0855.4+2009 & J0855.4+2009 & 0.31 & $4.81_{-0.09}^{+0.09}$ & $13.47_{-0.01}^{+0.01}$ & $-9.96_{-0.02}^{+0.02}$ & $21.40_{-0.09}^{+0.09}$ & $-10.50_{-0.04}^{+0.04}$ \\
\hline
J0538.8-4403 & J0538.8-4403 & 0.89 & $3.80_{-0.16}^{+0.19}$ & $13.27_{-0.03}^{+0.03}$ & $-10.64_{-0.02}^{+0.02}$ & $22.78_{-0.17}^{+0.25}$ & $-10.05_{-0.06}^{+0.10}$ \\
\hline
\end{tabular}}
\end{table}

\FloatBarrier
\begin{table}[h!]
\centering
\caption{Giommi BL Lac Table\label{giommi_bll_table}}
\rotatebox{0}{\begin{tabular}{|c|c|c|c|c|c|c|c|c|c|c|c|c|c|c|c|c|}
\hline
Source & SED Name & z & SF & $\text{Log}10(\nu_{s})$ & $\text{Log}10(\nu_{s}F_{\nu_{s}})$ & $\text{Log}10(\nu_{IC})$ & $\text{Log}10(\nu_{IC}F_{\nu_{IC}})$ \\
\hline
\hline
J0738+1742 & J0738+1742 & 0.42 & $3.03_{-0.20}^{+0.20}$ & $13.82_{-0.04}^{+0.04}$ & $-10.99_{-0.01}^{+0.01}$ & $23.75_{-0.19}^{+0.22}$ & $-11.10_{-0.09}^{+0.08}$ \\
\hline
J2202+4216 & J2202+4216 & 0.07 & $4.93_{-0.14}^{+0.14}$ & $13.75_{-0.04}^{+0.04}$ & $-10.29_{-0.02}^{+0.02}$ & $21.73_{-0.13}^{+0.13}$ & $-10.48_{-0.04}^{+0.04}$ \\
\hline
J0538-4405 & J0538-4405 & 0.89 & $3.69_{-0.20}^{+0.19}$ & $13.52_{-0.02}^{+0.02}$ & $-10.27_{-0.01}^{+0.01}$ & $23.01_{-0.19}^{+0.19}$ & $-9.92_{-0.03}^{+0.03}$ \\
\hline
J0818+4222 & J0818+4222 & 0.53 & $4.29_{-0.14}^{+0.14}$ & $13.09_{-0.03}^{+0.03}$ & $-11.45_{-0.02}^{+0.02}$ & $22.06_{-0.12}^{+0.12}$ & $-10.96_{-0.09}^{+0.09}$ \\
\hline
J0550-3216 & J0550-3216 & 0.07 & $8.63_{-0.08}^{+0.09}$ & $13.58_{-0.02}^{+0.02}$ & $-10.87_{-0.02}^{+0.02}$ & $18.11_{-0.08}^{+0.07}$ & $-10.57_{-0.03}^{+0.03}$ \\
\hline
J1800+7828 & J1800+7828 & 0.68 & $4.43_{-0.12}^{+0.12}$ & $13.59_{-0.02}^{+0.02}$ & $-10.82_{-0.01}^{+0.01}$ & $22.10_{-0.12}^{+0.12}$ & $-10.95_{-0.05}^{+0.05}$ \\
\hline
J1517-2422 & J1517-2422 & 0.05 & $4.76_{-0.08}^{+0.08}$ & $13.72_{-0.03}^{+0.03}$ & $-10.34_{-0.02}^{+0.02}$ & $21.79_{-0.08}^{+0.08}$ & $-10.70_{-0.03}^{+0.03}$ \\
\hline
\end{tabular}}
\end{table}

\FloatBarrier
\section{Supplementary Methods}\label{supp_methods}
\subsection{Empirical Model Fitting of SEDs}
    All SEDs were fit using the continuous simulated annealing global optimization algorithm \citep{CoranaACMTMS1987} as implemented by Goffe, Ferrier, and Rodgers \citep{GoffeJoE1994}, version 3.2.

    \subsubsection{Models} We used a few different empirical models to fit the SEDs. These are a log-parabola, log-parabola with a thermal (big-blue bump) component, third-degree polynomial, third-degree polynomial with a thermal (big-blue bump) component, and a log-parabola with a low-energy power-law cutoff (to model extended emission as necessary in a very small number of SEDs). Choosing between these models was based on $3\sigma $ significance preference over the null-hypothesis model (i.e., via Wilk's theorem imposing a $3\sigma $ cutoff to reject the model with less degrees of freedom). Model choice only disobeyed this criteria in a very small number of cases where the null-hypothesis model clearly visually disagreed with the SED data (in almost all such cases the significance of rejecting the null-hypothesis model was nearly $3\sigma $).

    We list the mathematical expressions for our models below. Here the different models are listed as LP (Log-Parabola), Poly$3$D (third-degree polynomial), BBB (big-blue bump thermal spectrum), LPBBB (log-parabola with big-blue bump added), Poly$3$DBBB (third-degree polynomial with big-blue bump added), PL (power-law), and LPPL (log-parabola with low-energy power-law cutoff). In Poly$3$D $\nu _{l}$ is the base-$10$ logarithm of the frequency, and $\nu_{l,0}$ is the pivot logarithmic frequency. BBB and PL are defined alone only to aid in comprehension. These two models were never used alone.

    \begin{equation}
        \text{LP}(\nu ) = N_{0}\left(\frac{\nu }{\nu _{p}}\right)^{-b\text{Log}10\left(\nu /\nu _{p}\right)}
    \end{equation}

    \begin{equation}
        \text{Poly}3\text{D}(\nu_{l}) = a(\nu_{l} -\nu _{l,0})^{3} + b(\nu_{l} -\nu _{l,0})^{2} + c(\nu_{l} -\nu _{l,0}) + d
    \end{equation}

    \begin{equation}
        \text{BBB}(\nu ) = \pi N_{\text{BBB}}\frac{2h\left(\nu \left(1+z\right)\right)^{4}}{c^{2}}\left(\text{Exp}\left[\frac{h\nu \left(1+z\right)}{k_{B}T_{\text{BBB}}}\right]-1\right)^{-1}
    \end{equation}

    \begin{equation}
        \text{LPBBB}(\nu ) = \text{LP}(\nu ) + \text{BBB}(\nu )
    \end{equation}

    \begin{equation}
        \text{Poly}3\text{DBBB}(\nu ) = \text{Poly}3\text{D}(\text{Log}10(\nu) ) + \text{BBB}(\nu )
    \end{equation}

    \begin{equation}
        \text{PL} = N_{0}\left(\frac{\nu }{\nu _{0}}\right)^{-\alpha }
    \end{equation}

    \begin{equation}
        \text{LPPL}(\nu) = \begin{cases}
            \text{LP}(\nu ), & \text{if } \nu >= \nu _{0}\\
            \text{PL}(\nu ), & \text{if } \nu < \nu _{0}
        \end{cases}
    \end{equation}

    We corrected data for attenuation due to the extragalactic background light using the Dominguez 2011 \citep{DominguezMNRAS2011} model as implemented in \verb|gammapy|.

\subsection{Effective Coverage of SEDs}\label{supp_methods:effective_coverage}
    Well-sampled in the context of our paper was defined using spectral curvature. SEDs were excluded if a power-law was preferred over a log-parabola at the $>=2\sigma $ level. We chose to use curvature as a proxy for ''well-sampledness`` since blazar spectral components are expected to have curvature. Therefore, if an SED cannot significantly constrain the curvature of either the synchrotron or IC component, the SED must not be well-sampled (either due to actual wavelength coverage or due to measurement uncertainties). A statistical test for what we term ''effective coverage`` is preferred over observer choice, since it reduces the possibility of any observer bias and places all SEDs on an equal statistical footing.

\subsection{Bootstrapping Methods}\label{boot_meth}
    \subsubsection{Error-Modified Bootstrap.} To estimate uncertainties of quantities for which the likelihood is unknown or unreasonable to calculate or sample, uncertainties can be estimated via bootstrapping. Bootstrapping is defined by estimating a population distribution by creating a sample through the use of resampling with replacement. Key to this discussion is the idea of bootstrapping as a method of sampling an observed distribution. It is upon this idea that we create an extension of the bootstrap to incorporate the uncertainties estimated for the observed variates (that is the observed quantities in a sample; e.g., the observed seed factors).

    The error-modified bootstrap extends the standard bootstrap by drawing, for each resampling of the observed variates, a random value of each individual resampled variate from its respective distribution as defined by its respective uncertainties. This creates a population distribution estimate which incorporates the measured sample uncertainties. This method also allows for an asymptotically continuous distribution of sample statistics (i.e., continuous in the limit of an infinite number of samples), as opposed to the necessarily discrete distribution produced by the standard bootstrap (since there is a finite number of combinations of observed variates in the case of the standard bootstrap).

    \subsubsection{Parameter Bootstrap.} For uncertainty estimation of values determined by a set of parameters (e.g., the peak frequency of a third-degree polynomial), it was necessary to perform a bootstrap on the measured parameters. This procedure is to create a random sample of parameters where each sample parameter, $\theta _{i}^{*}$, is drawn from the distribution of $\hat{\theta _{i}}$, as defined by the uncertainties on $\hat{\theta _{i}}$. That is, a single sample in the bootstrap is a set $\{\theta _{i}^{*}\}$ such that each $\theta _{i}^{*}$ has been drawn from the distribution of $\hat{\theta _{i}}$.

    This may not appear on the surface to be the same as the above-described error-modified bootstrap. However, this is a special case of the error-modified bootstrap. Here there is a single datum for the quantity in question. Since this quantity is a function of the given parameters, the distribution of the quantity is a function of the distributions of the parameters. Therefore this parameter bootstrap is really the error-modified bootstrap in the case of a single datum.

\subsection{Error Analysis}
    Error analysis (meaning here estimation of uncertainties) was accomplished through different means, depending on the application. Here we describe our methods for error analysis for SED parameters, seed factor parameters, and the various other distributions related to the seed factor as appropriate.

    \subsubsection{SED Error Analysis.} Error analysis of the parameters of the SED models was accomplished via application of Wilk's Theorem through the profile-likelihood method. For a given parameter, $\hat{\theta _{n}}$, for which the uncertainties were to be calculated, the log-likelihood of a given SED model was calculated for a sample of values, $\{\theta _{n}^{*}\}$, with all other parameters fixed at their best-fit values. This sample of log-likelihoods was then transformed by subtracting from all the log-likelihood values the difference of the best-fit log-likelihood and the target delta-log-likelihood (as given by Wilk's Theorem). This transformation was performed so that a root-finding method could be employed to find the uncertainties. This transformed sample was then interpolated using \verb|scipy.interpolate.splrep|, and \verb|scipy.interpolate.sproot| was used to find the roots.

    Error analysis of the peak frequencies and peak fluxes was performed differently depending on the SED model preferred for a given SED. In the case of log-parabolic models, the peak frequency and peak flux are parameters of the model. Therefore the uncertainties of the peak frequency and peak flux for any log-parabolic model was calculated using the profile-likelihood implementation described above.

    In the case of third-degree polynomials, however, the peak frequency and peak flux are not parameters of the model, and are instead calculated from the model after fitting. The uncertainties therefore cannot be calculated directly using the profile-likelihood method. A bootstrapping procedure (parameter bootstrap, as described in \ref{boot_meth}) was therefore employed to estimate the uncertainties. This method relied on first estimating the uncertainties of the model parameters through the profile-likelihood method, and then creating a large sample ($10^{6}$ for a pure third-degree polynomial and $10^{7}$ for a third-degree polynomial with added big-blue bump) of peak values calculated for parameter tuples drawn randomly from binormal distributions as defined by the parameter uncertainties calculated beforehand. A kernel-density estimate was then created from the bootstrapped peak values using \verb|sklearn.neighbors.KernelDensity| to provide an estimate of the log-likelihood function of each peak value. The bandwidth used for the kernel density estimate in this method was chosen based on testing. This log-likelihood estimate was then sampled and interpolated in the same way as for the model parameter error analysis, and roots were found to estimate the uncertainties of the peak values.

    \subsubsection{Seed Factor Error Analysis.} The uncertainties for the seed factor calculated for each individual SED were estimated using the error-modified bootstrap. The uncertainty on the seed factor population median was estimated using a bootstrap method (parameter bootstrap, as described in \ref{boot_meth}) on the individual seed factors. This was done by sampling the observed seed factor distribution with replacement, drawing, from each seed factor selected in a given sample, a value from the distribution defined by its uncertainties. In effect this is a modification of the bootstrap method which incorporates the uncertainties of the measured seed factors. $1\sigma $ confidence intervals were then constructed by finding the values on either side of the peak of the median seed factor distribution at which half the probability of a normal distribution $1\sigma $ interval was enclosed (that is to say that on the upper and lower sides of the median, it was found at what point, rounding for brevity to the nearest whole number, $34\%$ of the distribution was enclosed).

\subsection{Kernel Density Estimation of the Seed Factor Distribution}
    A kernel density estimate was calculated for a bootstrapped sample of the observed distribution of seed factors. Bootstrap samples were calculated using the seed factor uncertainties to better approximate the distribution. The bandwidth of the kernel density estimate was chosen using Silverman's Rule \citep{Silverman_density_estimation_textbook_1998}. Silverman's Rule gives a bandwidth which approximates the bandwidth which would minimize the mean integrated squared error. Silverman's Rule gives a bandwidth,

    \begin{equation}
        h = \left(\frac{4\sigma^{5}}{3n}\right)^{1/5}
    \end{equation}

    where $\sigma$ is the sample standard deviation, and $n$ is the number of data. In our method, the standard deviation of the bootstrapped distribution was taken as $\sigma$ (being an estimate of the population standard deviation), and the number of observed data was taken as $n$. We used a Gaussian kernel for the kernel density estimate.

\clearpage
\newcommand{\noop}[1]{}